\documentclass[12pt]{article} 
\pdfoutput=1
\usepackage[utf8]{inputenc}
\usepackage[T1]{fontenc}
\usepackage{lmodern, amsmath, amssymb, slashed, graphicx, comment,adjustbox}
\usepackage[dvipsnames]{xcolor}
\usepackage[numbers, elide, compress]{natbib}
\usepackage[colorlinks=true, citecolor=blue, urlcolor=blue, linkcolor=blue, breaklinks=true, pdfpagelabels=false]{hyperref}

\makeatletter\g@addto@macro\bfseries{\boldmath}\makeatother

\def\figureautorefname~#1\null{Fig.\,#1\null}

\def\equationautorefname~#1\null{Eq.\,(#1)\null}

\numberwithin{equation}{section}

\newcommand{\inab}{\,{\rm ab}^{-1}}

\newcommand{\eehz}{e^+e^- \to hZ}
\newcommand{\eevvh}{e^+e^- \to \nu \bar{\nu} h}
\newcommand{\eeww}{e^+e^- \to WW}

\newcommand{\vvh}{\nu \bar{\nu} h}

\newcommand{\eehzinv}{e^+e^- \to hZ,\,Z\to \nu\bar{\nu}}

\newcommand{\mrc}{m_{\rm recoil}}
\newcommand{\mrce}{m^E_{\rm recoil}}
\newcommand{\mrcp}{m^p_{\rm recoil}}

\newcommand{\La}{\mathcal{L} }

\newcommand{\bpm}{\begin{pmatrix}}
\newcommand{\epm}{\end{pmatrix}}

\newcommand{\hlb}[1]{{\color{blue} #1}}

\begin{document}

\begin{flushright}
DESY 17-129
\end{flushright}

\vspace*{2cm}

\begin{center}

{\large\bf
Optimizing Higgs factories by modifying the recoil mass
\par}
\vspace{9mm}

{\bf  Jiayin~Gu$^{a,b}$ and Ying-Ying Li$\, ^{c}$}\\ [4mm]
{\small\it
$^a$ DESY, Notkestra{\ss}e 85, D-22607 Hamburg, Germany \\[2mm]
$^b$ Center for Future High Energy Physics, Institute of High Energy Physics, \\ 19B YuquanLu, Chinese Academy of Sciences, Beijing 100049, China \\[2mm]
$^c$ Department of Physics, The Hong Kong University of Science and Technology,\\
Clear Water Bay, Kowloon, Hong Kong S.A.R., P.R.C
\par}
\vspace{.5cm}
\centerline{\tt \small jiayin.gu@desy.de, ylict@connect.ust.hk}

\end{center}

\begin{abstract}
It is difficult to measure the $WW$-fusion Higgs production process ($\eevvh$) at a lepton collider with a center of mass energy of 240-250\,GeV due to its small rate and the large background from the Higgsstrahlung process with an invisible~$Z$ ($\eehzinv$). We construct a modified recoil mass variable, $\mrcp$, defined using only the 3-momentum of the reconstructed Higgs particle, and show that it can better separate the $WW$-fusion and Higgsstrahlung events than the original recoil mass variable $\mrc$.  Consequently, the $\mrcp$ variable can be used to improve the overall precisions of the extracted Higgs couplings, in both the conventional framework and the effective-field-theory framework.  We also explore the application of the $\mrcp$ variable in the inclusive cross section measurements of the Higgsstrahlung process, while a quantitive analysis is left for future studies.  
\end{abstract}

\newpage
{\small 
\tableofcontents}

\setcounter{footnote}{0}

\vspace{2cm}
\section{Introduction}

A lepton collider running at a center of mass energy of around 240 to 250 GeV is ideal for studying the properties of the Higgs boson.   It can collect a large amount of Higgsstrahlung ($\eehz$) events, which has a cross section maximized at around 250\,GeV.  At higher energies, the $WW$-fusion process of Higgs production ($\eevvh$) can be better measured as its cross section increases with energy.  It is important to have good measurements of the $WW$-fusion process, which provides information complementary to the one from the Higgsstrahlung process.  In the conventional \emph{kappa} framework, the $WW$-fusion process can constrain the $hWW$ coupling and is also an important input for the determination of the Higgs total width.\footnote{See {\it e.g.} Ref.~\cite{Lafaye:2017kgf} for a recent study under this framework.}  In the effective-field-theory (EFT) framework, the $WW$-fusion and the Higgsstrahlung processes probe different combinations of EFT parameters.  The inclusion of both processes, as well as the diboson one ($\eeww$), is crucial for discriminating different EFT parameters and obtaining robust constraints on all of them~\cite{Ellis:2015sca, Durieux:2017rsg, Barklow:2017suo}.  However, it is not guaranteed that the runs at energies higher than 240-250\,GeV will be available.  The proposed Circular Electron Positron Collider (CEPC) in China does not have plans for the 350\,GeV run at the current moment~\cite{CEPC-SPPCStudyGroup:2015csa}.  For the Future Circular Collider (FCC)-ee at CERN~\cite{FCCupdate} and the International Linear Collider (ILC) in Japan~\cite{Baer:2013cma}, a significant amount of time may also be spent at the 240\,GeV/250\,GeV run before moving on to higher energies.  The measurements of the $WW$-fusion process at 240-250\,GeV is therefore of great relevance to the study of Higgs physics.

\begin{figure}[t]
\centering \hspace{1cm}
\includegraphics[height=2.6cm]{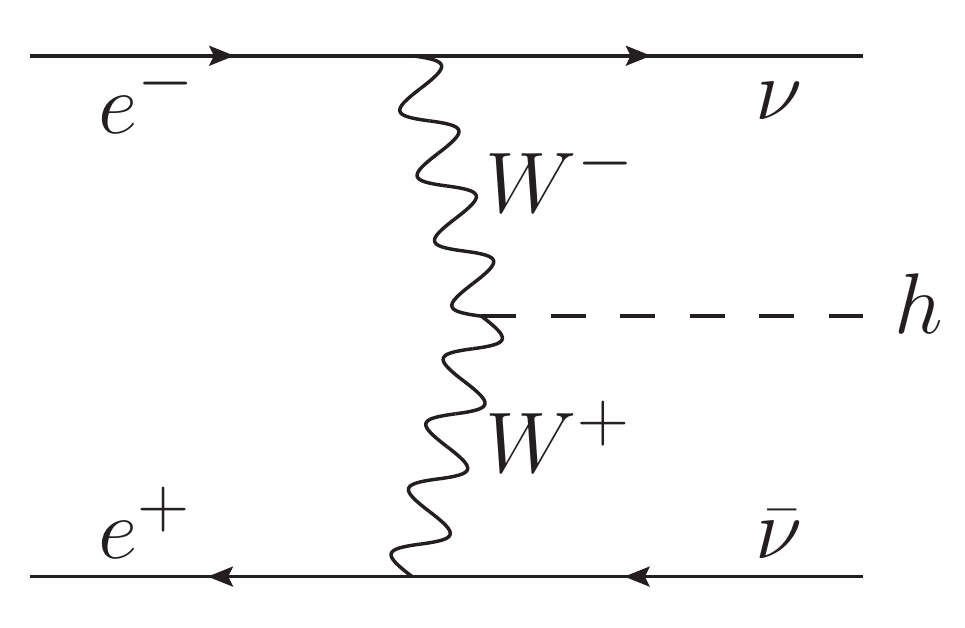}  \hspace{2cm}
\includegraphics[height=3cm]{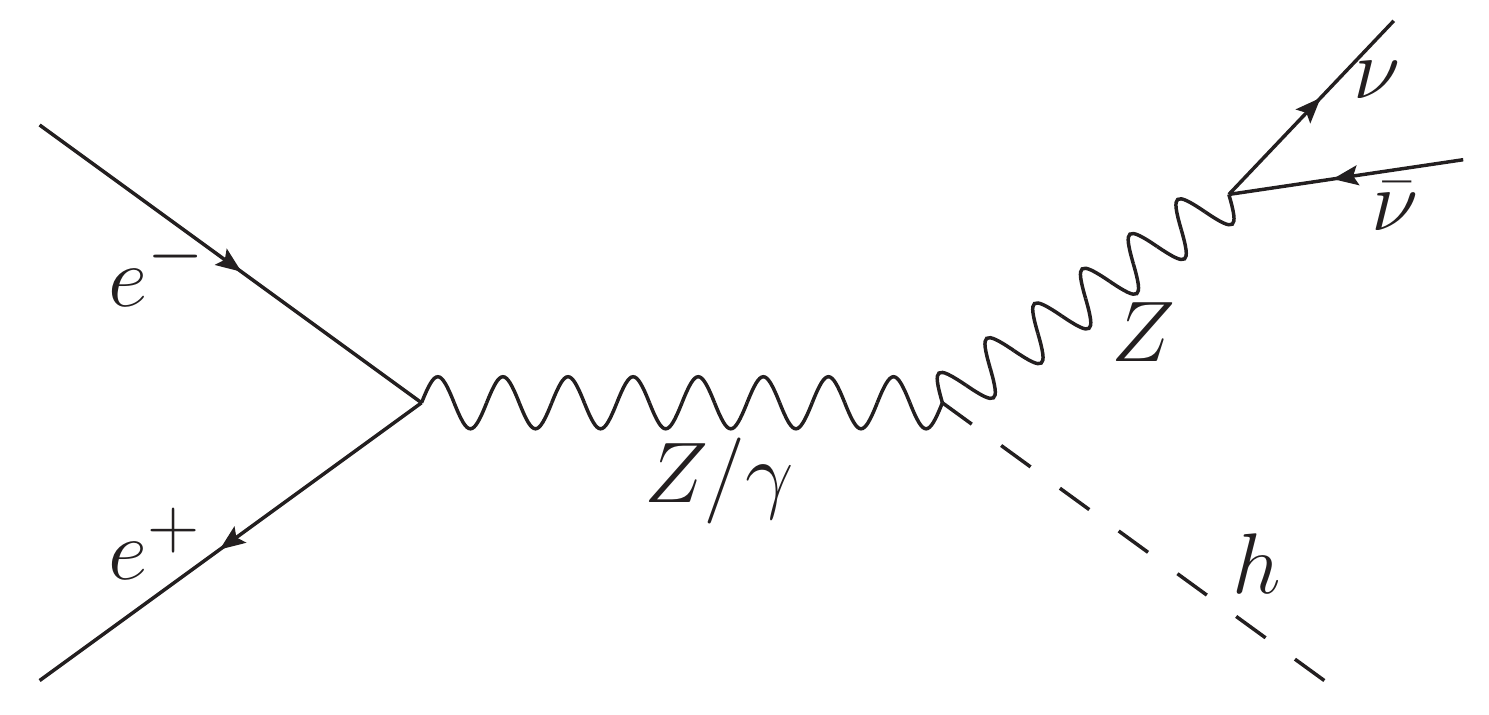}
\caption{The two processes that contributes to the $\eevvh$ channel.  {\bf Left:} The $WW$-fusion process of Higgs production.  {\bf Right:} The Higgsstrahlung process with $Z$ decaying to a pair of neutrinos.}
\label{fig:feynvvh}
\end{figure}

It is difficult to measure the $WW$-fusion process at 240-250\,GeV for the following two reasons.  First, it has a small rate at lower energy, with a cross section of 6.72\,fb at 250\,GeV assuming unpolarized beams, while the total cross section of the Higgsstrahlung process is 212\,fb at the same energy~\cite{CEPC-SPPCStudyGroup:2015csa}.  Second, the Higgsstrahlung process with $Z$ decaying invisibly ($\eehzinv$)  
is the dominate background of $WW$ fusion.  Both contributing to the channel $\eevvh$ as shown in \autoref{fig:feynvvh}, the cross section of the former is more than six times of the latter one.  With longitudinal beam polarizations, the situation is slightly better.  If the signs of the polarizations are ${\rm sgn}(P(e^-), P(e^+)) = (-, +)$, the $WW$-fusion cross section is enhanced, and by a larger factor than the one of $\eehz$.  The method of recoil mass can also be used to separate the $WW$-fusion and $hZ$ events, as the reconstructed mass of the neutrino pair should center around the $Z$ mass for $hZ$ events. 
However, the discriminating power is limited by the detector resolutions (especially for hadronic Higgs decays) and other effects~\cite{CEPC-SPPCStudyGroup:2015csa, Asner:2013psa, Durig:2014lfa}.   A consistent treatment of the $hZZ$ and $hWW$ couplings is also required since they are related by gauge invariance.  In the EFT framework, the relation is complicated by the inclusion of Dimension-6 operators which generate anomalous couplings with Lorentz structures different from the standard model (SM) ones.

In this paper, we try to address the issues mentioned above and further optimize the measurements in $\eevvh$ at 240-250\,GeV.  We first perform a collider study in \autoref{sec:mrecoil} with a comparison of the recoil mass variable and its variations. We try to validate our study by following closely Ref.~\cite{Durig:2014lfa}.  We point out that the variable $\mrcp$, defined using only the 3-momentum of the reconstructed Higgs particle, could provide a discriminating power  
better than the original recoil mass variable $\mrc$ does.   We then implement the $\mrc$ and $\mrcp$ distributions in the EFT global analysis  in \autoref{sec:heft}, using the framework in Ref.~\cite{Durieux:2017rsg}.  We point out the importance of fitting the EFT parameters directly to the binned $\mrc$ or $\mrcp$ distribution instead of fitting them to the extracted cross sections of the $WW$-fusion process.  
We also apply $\mrcp$ to the inclusive $\eehz$ process in \autoref{sec:hz} and comment on its potential use in the inclusive $hZ$ cross section measurements.  Finally, we conclude in \autoref{sec:con}.  We provide a short summary of the EFT framework used in our analysis in \autoref{app:eft} and the numerical expressions of the EFT dependence of the (modified) recoil mass distributions in \autoref{app:expression}.

The following collider scenarios are considered in our study:
\begin{itemize}

    \item  {\bf CEPC} with $5\inab$ data collected at 240\,GeV, with unpolarized beams \cite{CEPC-SPPCStudyGroup:2015csa}.    This scenario can also be thought as the earlier stage of the FCC-ee, which also plans to collect $5\inab$ at 240\,GeV and eventually $1.5\inab$ data at 350\,GeV as well \cite{FCCupdate}.
    
    \item   {\bf ILC} with $2\inab$ data collected at 250\,GeV and beam polarizations of $P(e^-, e^+) = (\pm0.8,\pm0.3)$, which could be considered as the first stage of a full program with center of mass energies up to 500\,GeV \cite{Barklow:2015tja}.

\end{itemize}
For the $WW$-fusion measurements, we focus on the channel with the Higgs decaying to a pair of bottom quarks ($\eevvh, h\to b\bar{b}$), which has the largest branching ratio.   The measurements of $WW$ fusion at 240-250\,GeV with other Higgs decay channels are not reported in the official documents   due to the poor constraints (see {\it e.g.}, Refs.~\cite{CEPC-SPPCStudyGroup:2015csa, Barklow:2015tja, Gomez-Ceballos:2013zzn}).


\section{The modified recoil mass of $\eevvh, h\to b\bar{b}$}
\label{sec:mrecoil}

At lepton colliders, the method of recoil mass can be used to reconstruct the mass of a particle without measuring its decay products.  One of its most important applications is the measurement of the inclusive rate of the Higgsstrahlung process, $\eehz$.  Assuming both the Higgs and $Z$ are on mass shell, one could write the relation
\begin{equation}
m^2_h = E_h^2 -   |\vec{p}_h|^2 = (\sqrt{s}-E_Z)^2 - |\vec{p}_Z|^2 \,,  \label{eq:mhrecoil0}
\end{equation}
where the total center of mass energy $\sqrt{s}$ is fixed up to corrections from  
beam energy spread and initial state radiations.  By measuring the energies and momenta of the $Z$ decaying products one could reconstruct the mass of the Higgs particle.  This can be used to select $\eehz$ signal events without tagging the Higgs decay products, which makes it possible to measure the inclusive cross section of this channel.  It also provides the best Higgs mass measurement.  For example, a precision of $5.9$\,MeV can be achieved with the leptonic $Z$ decay channels of the inclusive $hZ$ measurements at the CEPC~\cite{CEPC-SPPCStudyGroup:2015csa}.

If the Higgs decay products are measured, the recoil mass can be turned around to reconstruct the $Z$ mass, since the following relation also holds for an $\eehz$ event,
\begin{equation}
m^2_Z = (\sqrt{s}-E_h)^2 - |\vec{p}_h|^2 = s-2 \sqrt{s} \, E_h + m^2_h \,.
\end{equation}
The recoil mass can then be defined as
\begin{equation}
\mrc^2 = s-2 \sqrt{s}\, E^{\rm rec}_h + (m^{\rm rec}_h)^2 \,,  \label{eq:mrc}
\end{equation}
where $E^{\rm rec}_h$ and $m^{\rm rec}_h$ are the reconstructed Higgs energy and mass.  
For Higgs decaying to a pair of bottom quarks, they are the total energy and the invariant mass of the two $b$-jets. 
This offers a way to separate the Higgsstrahlung events with an invisible $Z$ ($\eehzinv$) from the $WW$-fusion events, both contributing to the channel $\eevvh$, as shown in \autoref{fig:feynvvh}.  However, due to finite jet resolutions, beam energy spread and other effects, the recoil mass distribution of the $hZ$ events has a rather large spread.  This limits it discriminating power especially at the energy 240-250\,GeV, for which the recoil mass distribution of the $WW$-fusion events spreads around the same region. We also find that for Higgs decaying to a pair of $b$-jets ($h\to b\bar{b}$) the uncertainty on the recoil mass is dominated by the energy and momentum resolutions of the $b$-jets.  This is also obvious from the observation that the recoil mass distribution for the Higgs mass reconstruction in \autoref{eq:mhrecoil0} 
is much narrower for the  leptonic $Z$ decay channel than for the hadronic one (see {\it e.g.} Ref.~\cite{CEPC-SPPCStudyGroup:2015csa}). 

\begin{figure}[t]
\centering
\includegraphics[width=0.46\textwidth]{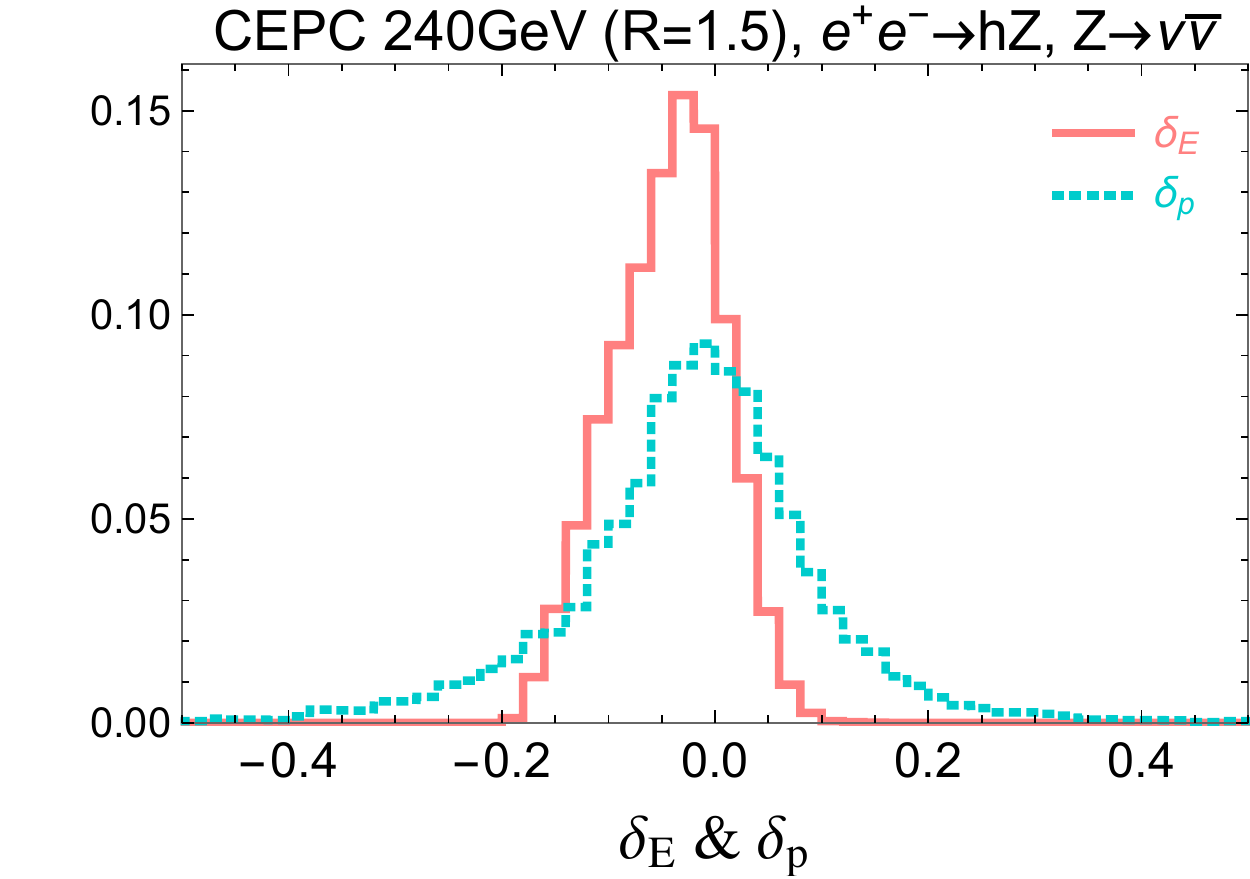} \hspace{0.1cm}
\includegraphics[width=0.48\textwidth]{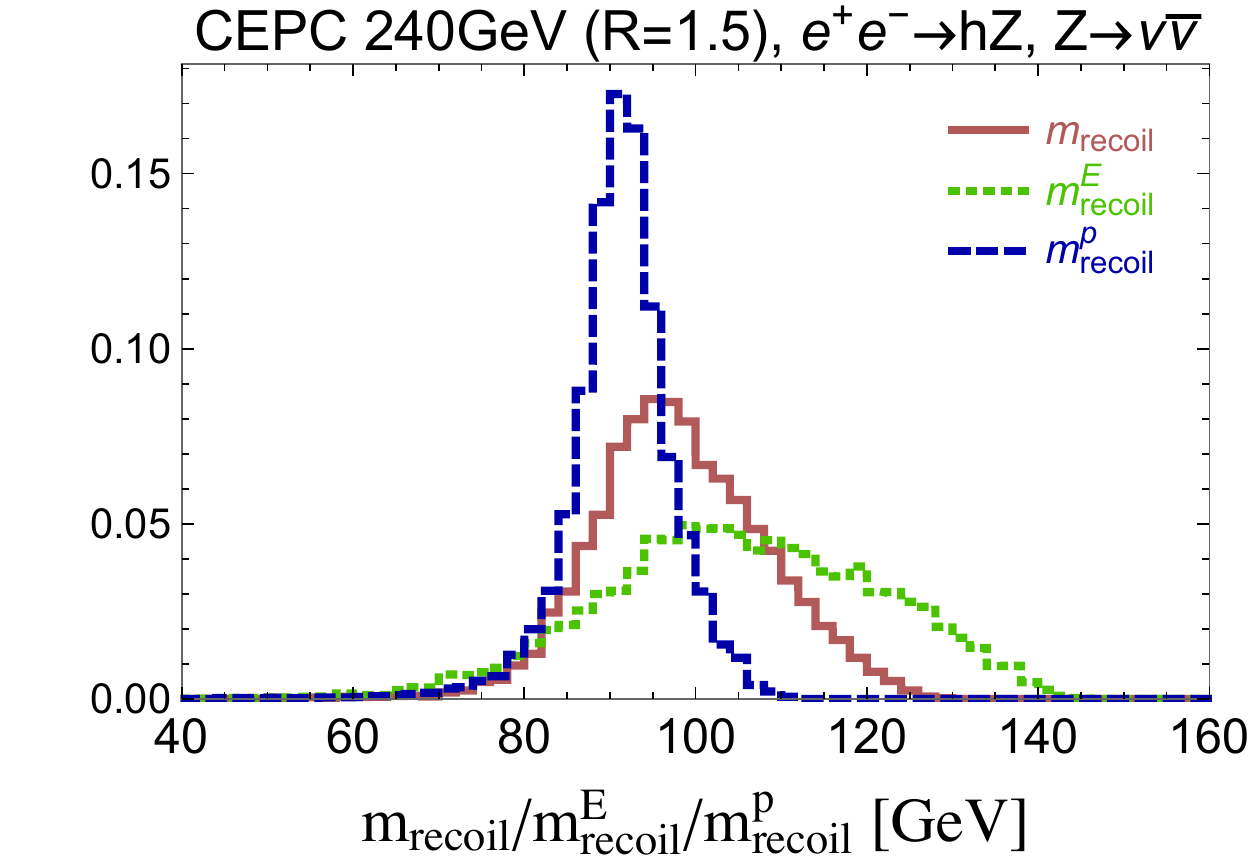}   
\caption{ {\bf Left:} The distributions of $\delta_E$ and $\delta_p$ (defined in \autoref{eq:dep}) for $\eehzinv, h\to b\bar{b}$ at CEPC 240\,GeV, after applying a Higgs-mass-window cut of $105\,{\rm GeV}<m^{\rm rec}_h<135\,{\rm GeV}$.  {\bf Right:} The corresponding distributions of $\mrc$, $\mrce$ and $\mrcp$, defined in \hlb{Eqs.\,(\ref{eq:mrc}\,--\,\ref{eq:mrcp})}.   A radius of $R=1.5$ is used in the jet clustering algorithm. The details on the simulations and selection cuts are stated later in this section.}
\label{fig:depm1}
\end{figure}

\begin{figure}[t]
\centering
\includegraphics[width=0.45\textwidth]{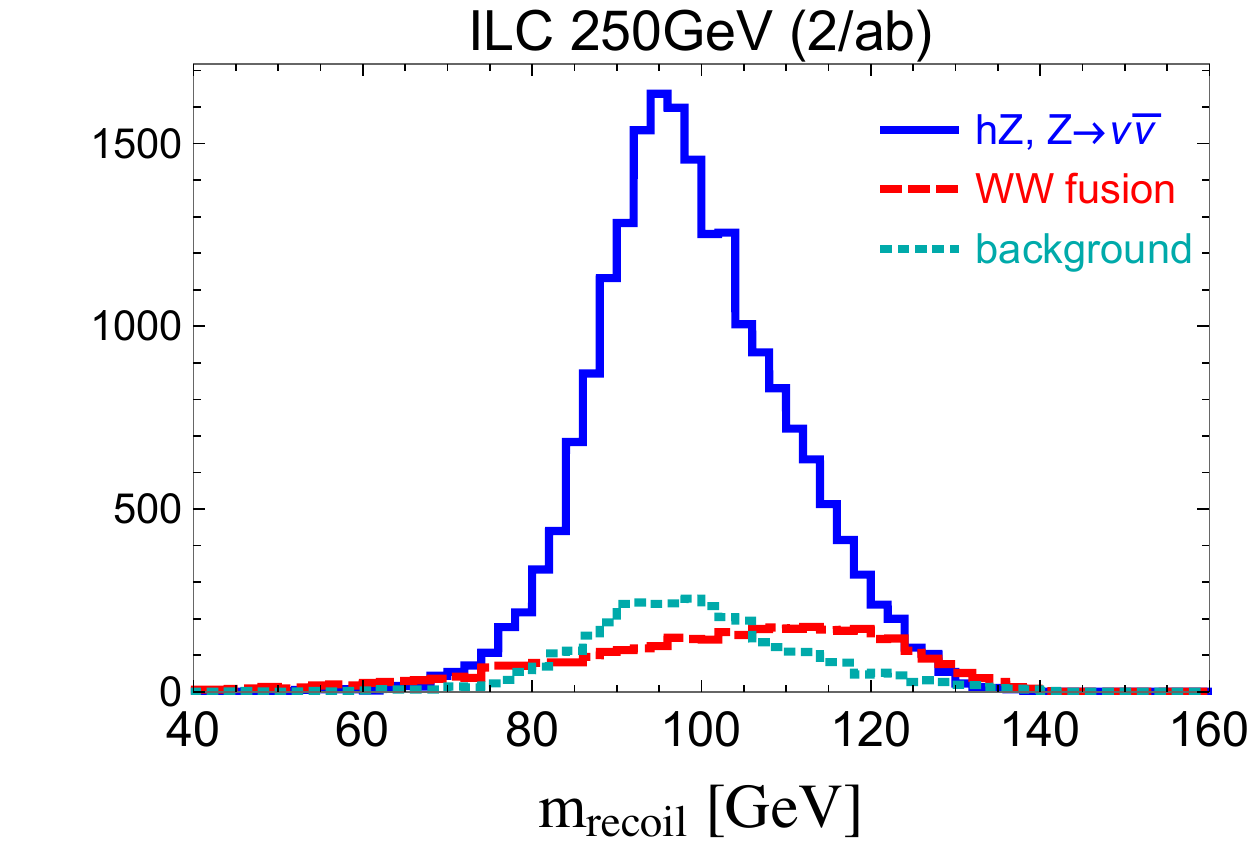} \hspace{0.3cm}
\includegraphics[width=0.45\textwidth]{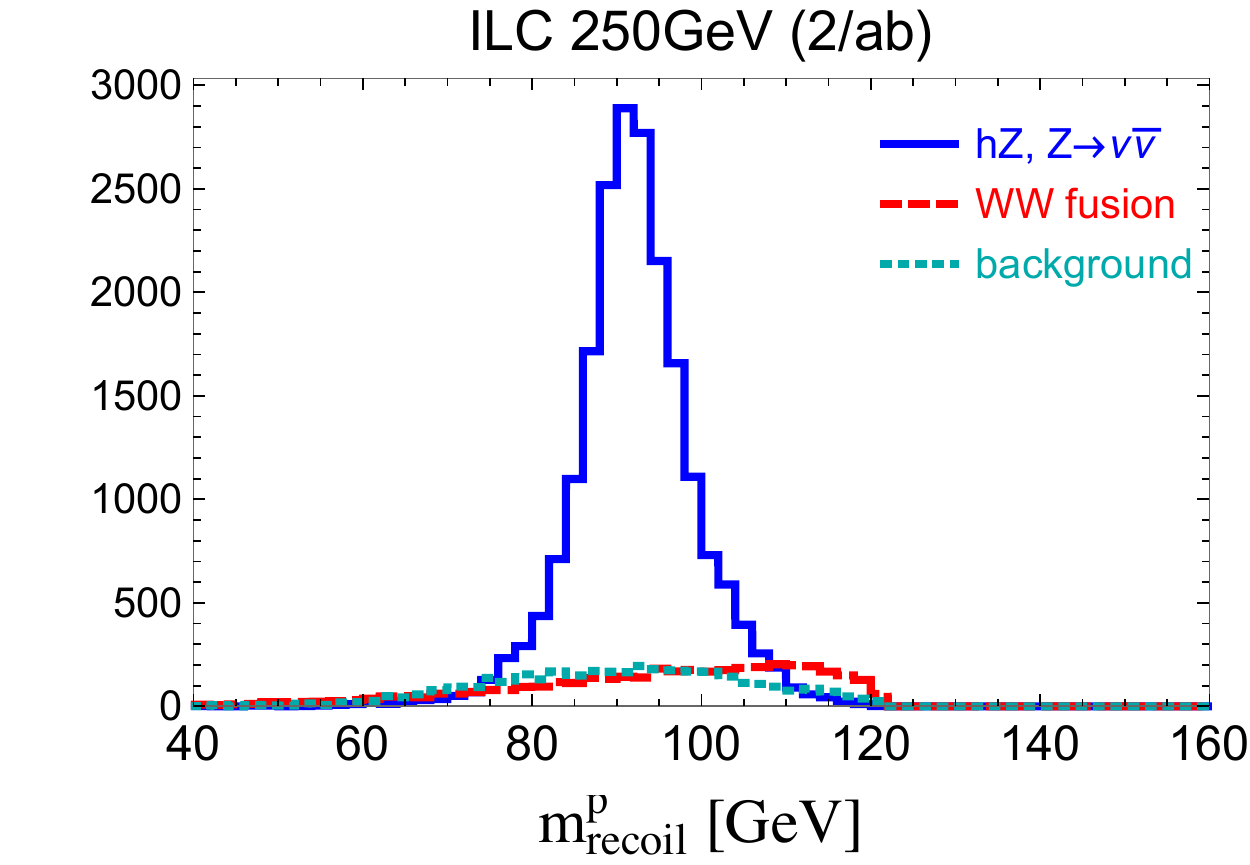}  \\ \vspace{0.3cm}
\includegraphics[width=0.45\textwidth]{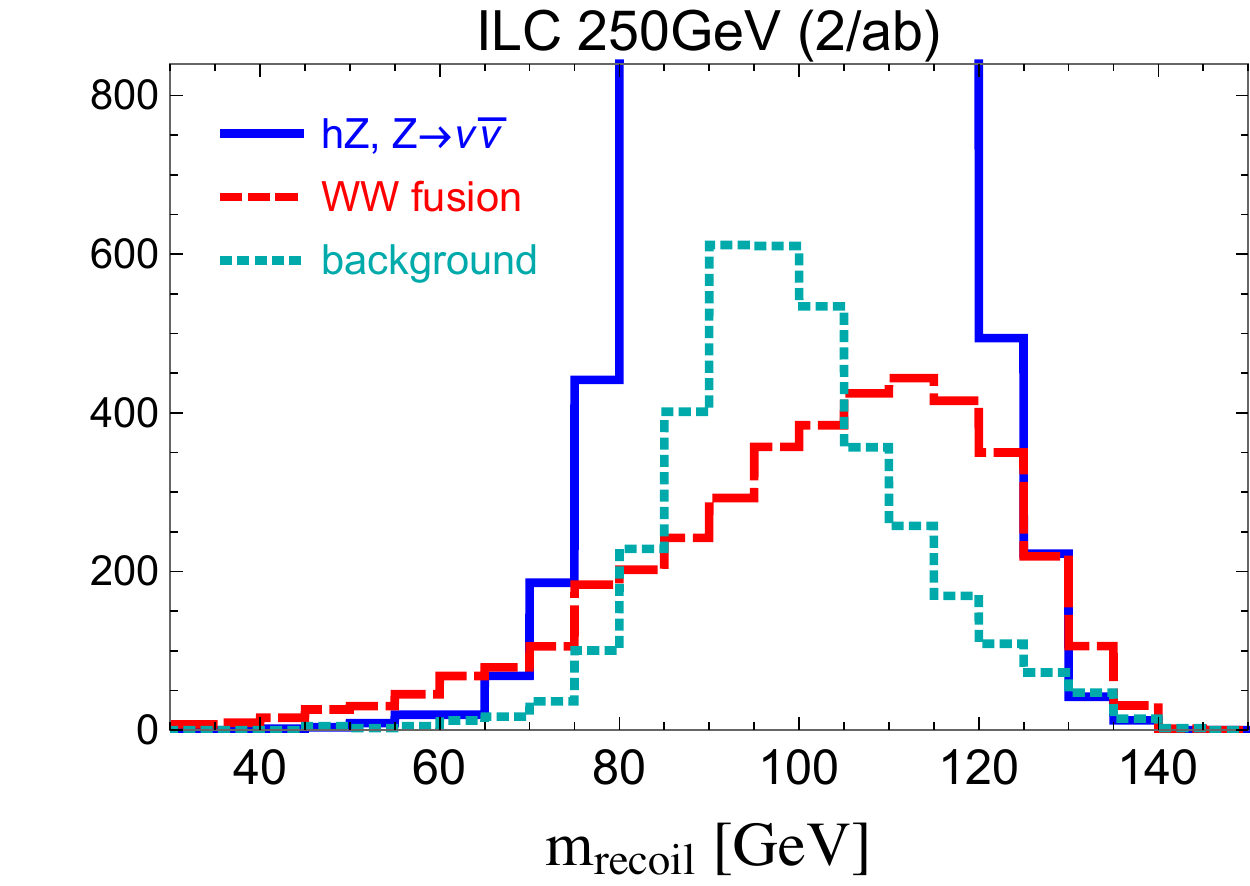} \hspace{0.3cm}
\includegraphics[width=0.45\textwidth]{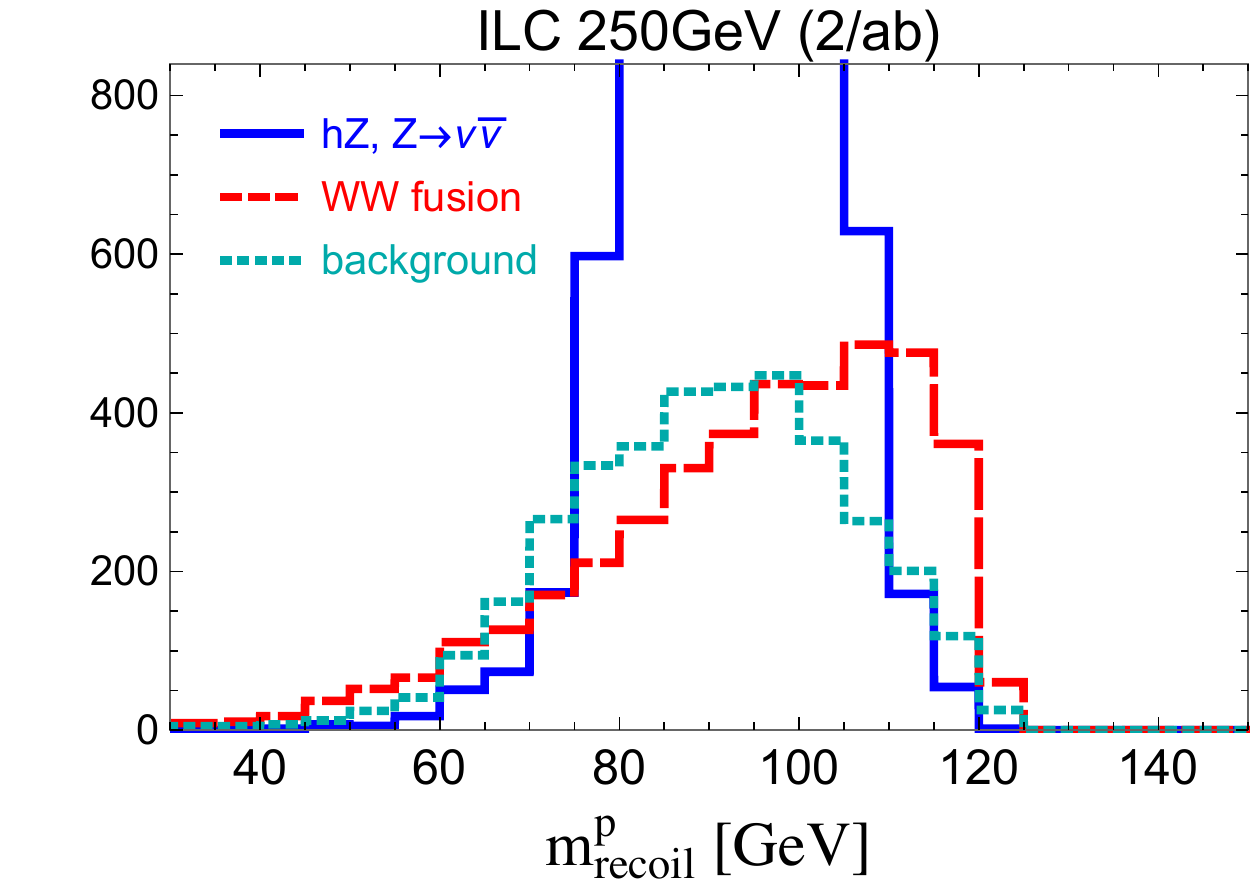}
\caption{The $\mrc$ (left panel) and $\mrcp$ (right panel) distributions of $hZ$ ($Z\to \nu\bar{\nu}$), $WW$-fusion and background events after selection cuts for ILC 250\,GeV with a luminosity of $2\inab$ and beam polarization $P(e^-,e^+) = (-0.8,+0.3)$. The distributions in the bottom panels are the amplified versions of the ones in the top panels (and also with a different bin size). }
\label{fig:nfitilc}
\end{figure}

\begin{figure}[t]
\centering
\includegraphics[width=0.45\textwidth]{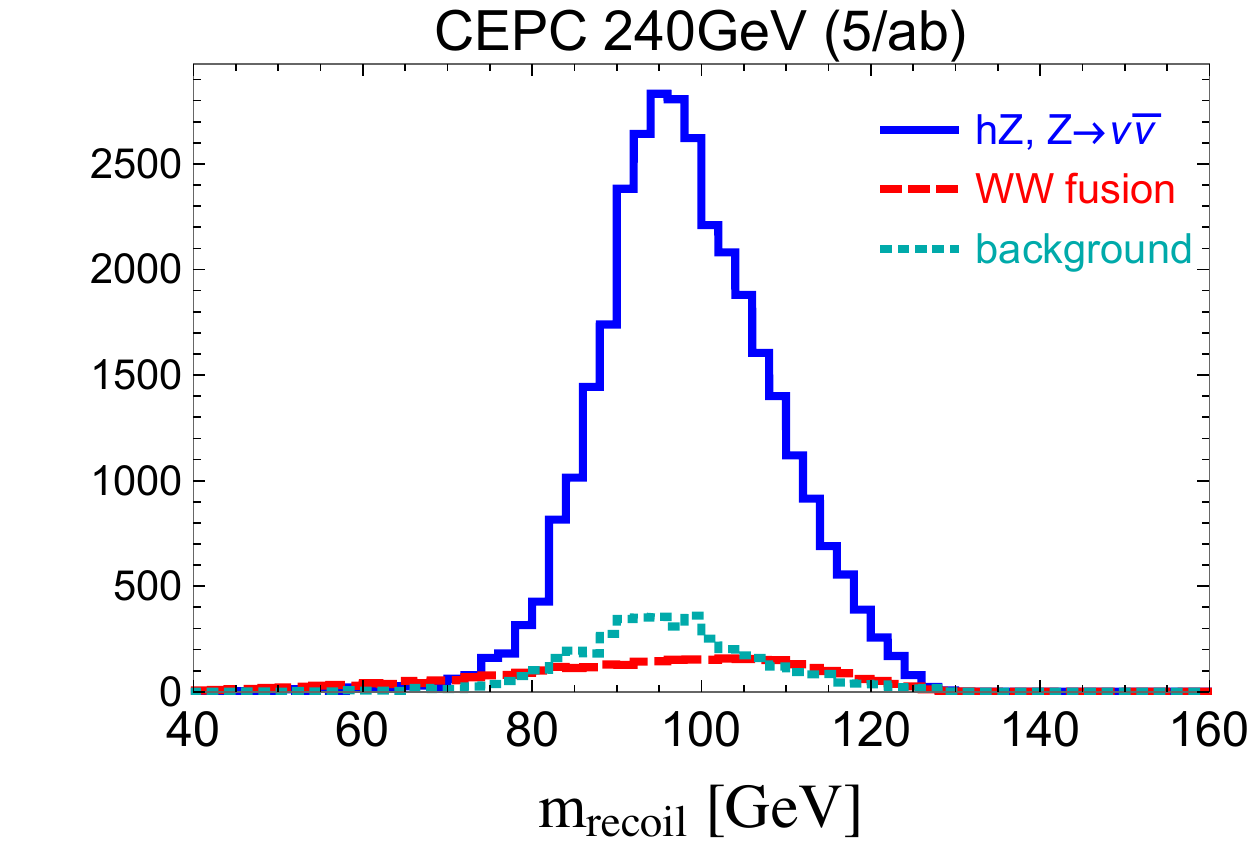} \hspace{0.3cm}
\includegraphics[width=0.45\textwidth]{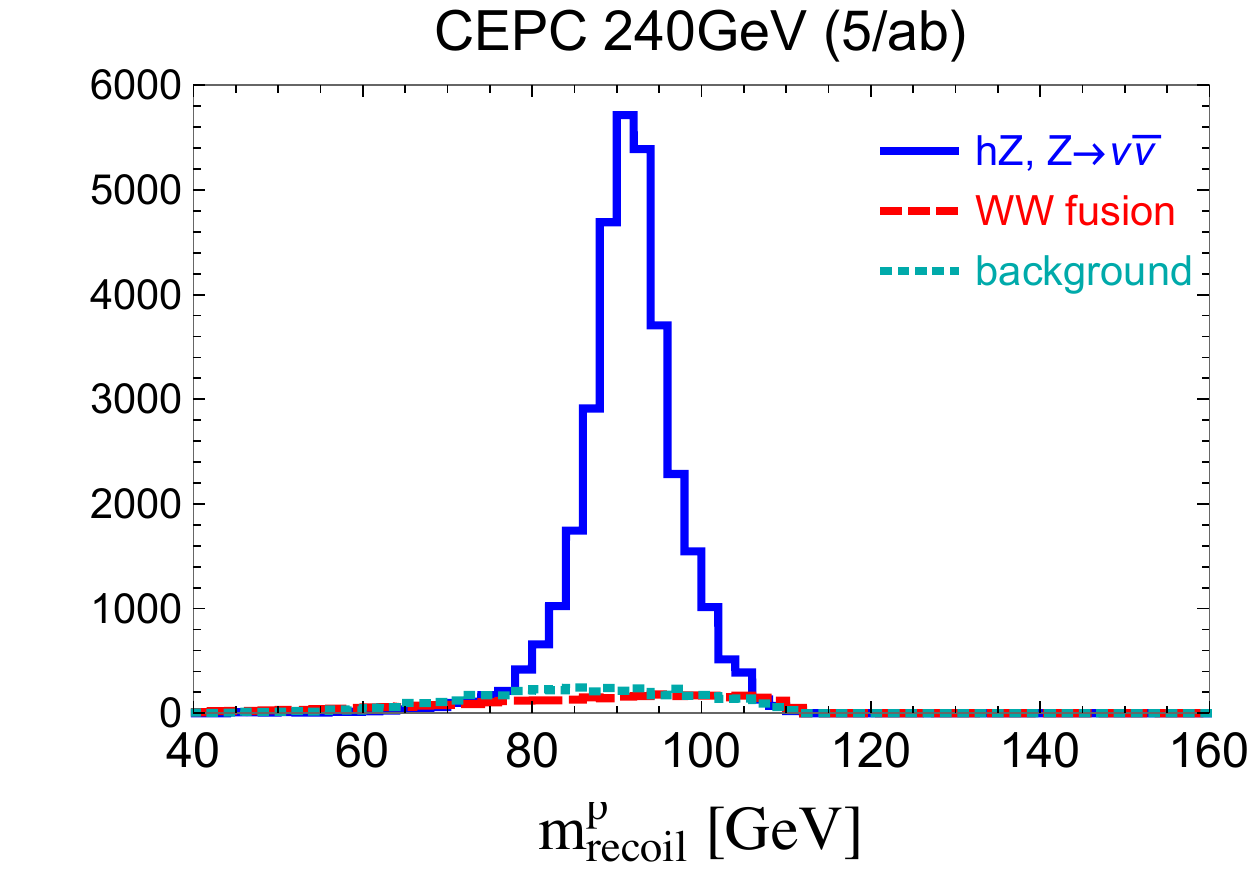}  \\ \vspace{0.3cm}
\includegraphics[width=0.45\textwidth]{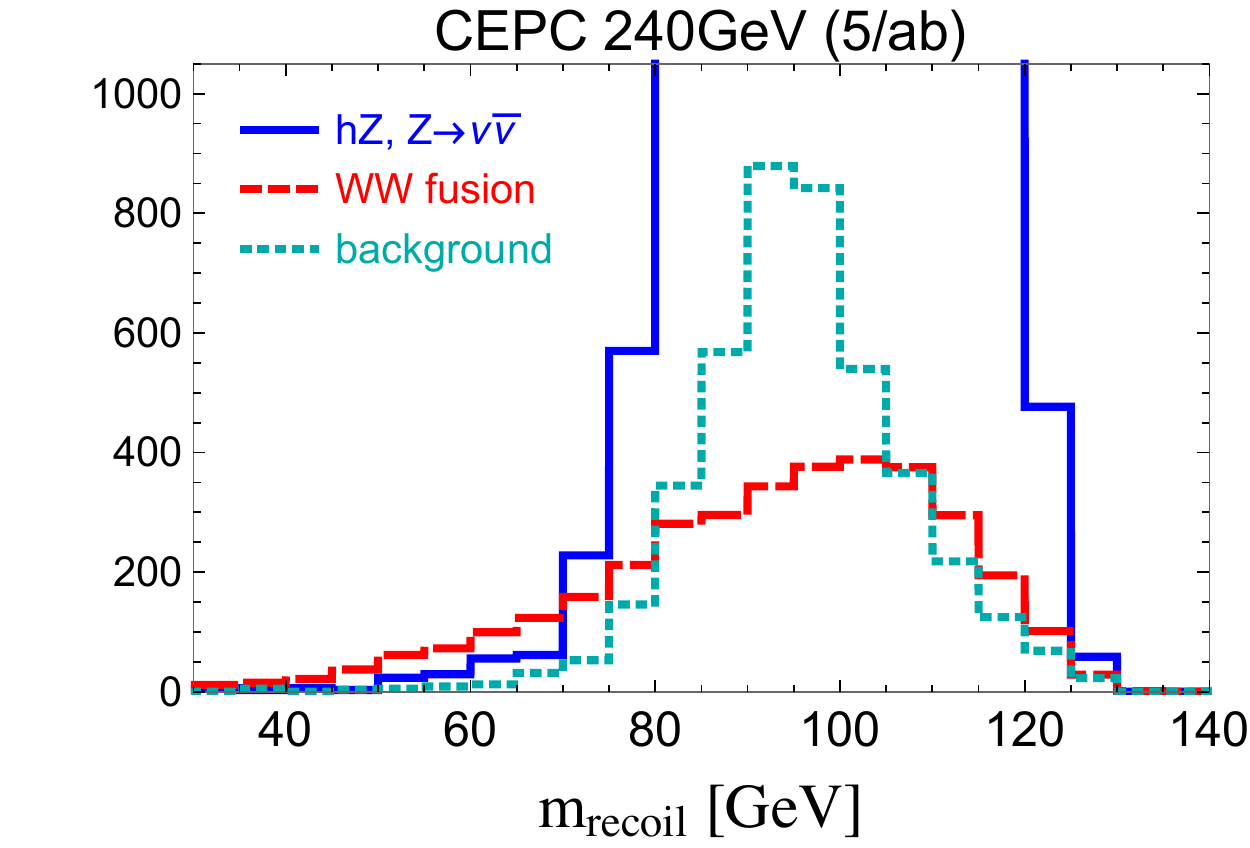} \hspace{0.3cm}
\includegraphics[width=0.45\textwidth]{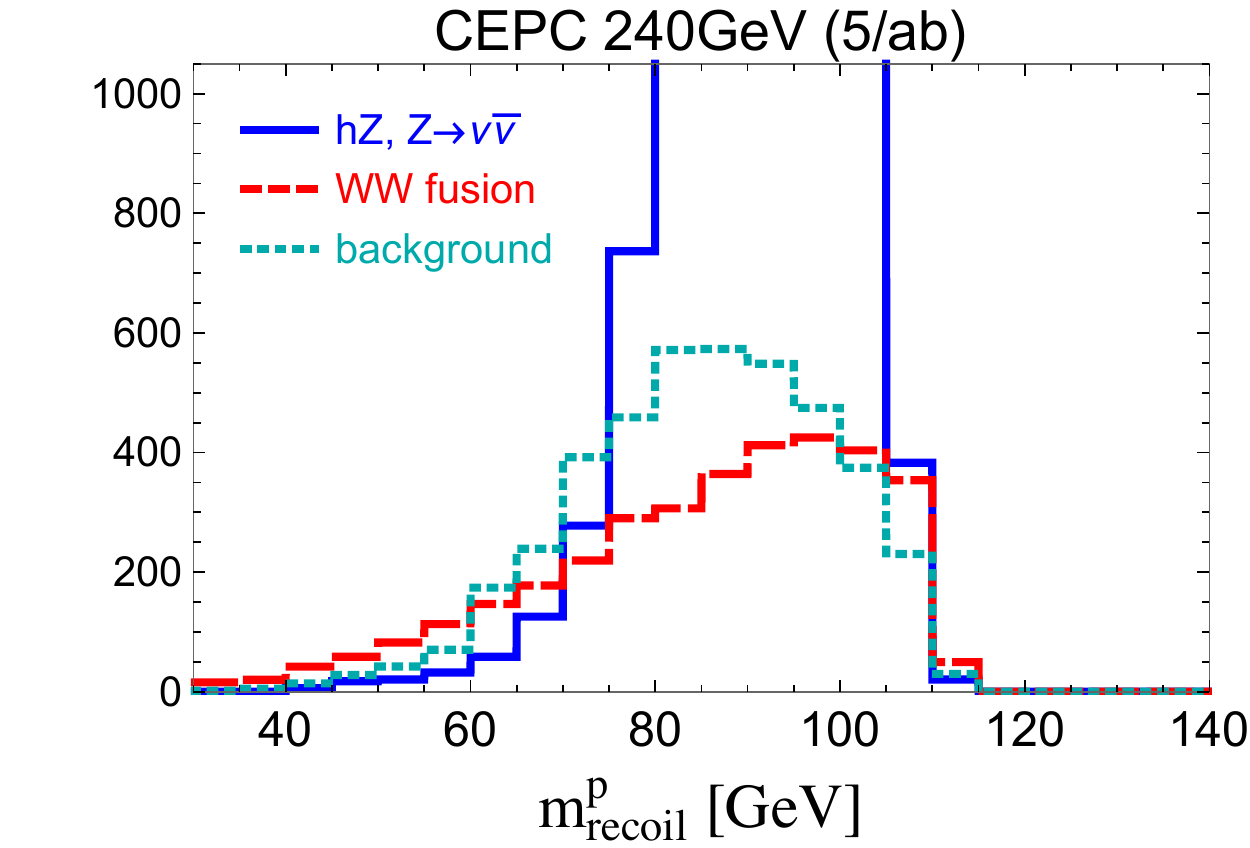}
\caption{Same as \autoref{fig:nfitilc} but for CEPC 240\,GeV with $5\inab$ data and unpolarized beams.}
\label{fig:nfitcepc}
\end{figure}

While the recoil mass defined in \autoref{eq:mrc} makes use of all the kinematic information, it does not make any assumption on the value of the Higgs mass.  Both the Higgs width and the projected uncertainty of its mass are at the MeV level and can be neglected compared with the effects of jet resolution.  Using the information of the Higgs mass, two modifications of the recoil mass can be constructed.  The first, using only the reconstructed Higgs energy as the measurement input, is defined as
\begin{equation}
(\mrce)^2 = s-2 \sqrt{s}\, E^{\rm rec}_h + m^2_h \,,  \label{eq:mrce}
\end{equation}
where $m_h$ is fixed to be the Higgs mass, $125.09$\,GeV.  The other, using only the reconstructed 3-momentum ($\vec{p}_h^{\rm\, rec}$) of the Higgs, is defined as
\begin{equation}
(\mrcp)^2 = s-2 \sqrt{s}\, \sqrt{m^2_h+|\vec{p}_h^{\rm \, rec}|^2} + m^2_h \,,  \label{eq:mrcp}
\end{equation}
where $m_h$ is again fixed to be $125.09$\,GeV.  At the truth level, $\mrc$, $\mrce$ and $\mrcp$ are all equivalent.  However, the uncertainties in the energy and momentum measurements certainty have different impacts on the three variables.  
To illustrate this impact, we define, for a given event, a set of five parameters \{$\delta_m$, $\delta_m^E$, $\delta_m^p$, $\delta_E$, $\delta_p$\} which parameterize the differences between the reconstructed quantities and the true ones, with
\begin{equation}
\mrc =~ \mrc^{\rm true} (1+\delta_m) \,,~~~~  \mrce = \mrc^{\rm true} (1+\delta_m^E) \,,~~~~   \mrcp = \mrc^{\rm true} (1+\delta_m^p) \,,  \label{eq:dmmm}
\end{equation}
and
\begin{equation}
E^{\rm rec}_h  =~ E_h (1+\delta_E) \,,~~~~~~~~~~~   |\vec{p}_h^{\rm \, rec}| = |\vec{p}_h| (1+\delta_p) \,,  \label{eq:dep}
\end{equation}
where $\mrc^{\rm true}$ is the true parton level recoil mass, $E_h$ and $\vec{p}_h$ are the true energy and 3-momentum of the Higgs.  For $hZ$ events, $\mrc^{\rm true}= m_Z$ (assuming it is on shell), and the three parameters $\delta_m$, $\delta_m^E$ and $\delta_m^p$ can be written in terms of $\delta_E$ and $\delta_p$.  At leading order, they are given (for $hZ$ events) by
\begin{align}
\delta_m \approx&~ -\frac{1}{m^2_Z} \left[ (\sqrt{s}-E_h)\, E_h \, \delta_E + |\vec{p}_h|^2 \, \delta_p \right] \,,  \nonumber\\
\delta_m^E \approx&~ -\frac{\sqrt{s}}{m^2_Z}  \, E_h \, \delta_E  \,,  \nonumber\\
\delta_m^p \approx&~ -\frac{\sqrt{s}}{m^2_Z} \,  \frac{ |\vec{p}_h|^2}{E_h} \, \delta_p \,.  \label{eq:dmzana}
\end{align}
Note that $\delta_E$ and $\delta_p$ can be either positive or negative.  The overall negative coefficients in \autoref{eq:dmzana} indicates that if the measured energy or 3-momentum of the Higgs is larger than its actual value, the recoil mass variables will be smaller than the $Z$ mass, and {\it vice versa}.  For a fixed center of mass energy ($\sqrt{s}=240$\,GeV or 250\,GeV), the values of $E_h$ and $|\vec{p}_h|$ are fixed.  In particular, near the $hZ$ threshold $|\vec{p}_h|$ is significantly smaller than $E_h$. With $|\vec{p}_h| \approx 51$\,GeV and $E_h\approx135$\,GeV at $\sqrt{s}=240$\,GeV, and $|\vec{p}_h| \approx 62$\,GeV and $E_h\approx140$\,GeV at $\sqrt{s}=250$\,GeV, 
\autoref{eq:dmzana} thus becomes
\begin{equation}
\delta_m ~/~ \delta_m^E ~/~ \delta_m^p \approx  
\left\{ \begin{matrix} 
-1.7 \, \delta_E -0.32 \, \delta_p ~/~ -3.9 \, \delta_E ~/~  -0.57 \, \delta_p \hspace{1cm}  \mbox{ at } 240\,{\rm GeV} \\  
-1.9 \, \delta_E -0.46 \, \delta_p ~/~  -4.2 \, \delta_E ~/~  -0.83 \, \delta_p  \hspace{1cm}  \mbox{ at } 250\,{\rm GeV} 
\end{matrix} \right. \,, \label{eq:dmz}
\end{equation}
where the small coefficients of $\delta_p$ come from a suppression factor of $\sim |\vec{p}_h|^2/E^2_h$ relative to the ones of $\delta_E$, shown in \autoref{eq:dmzana}.  
The distributions of $\delta_E$ and $\delta_p$ for $\eehzinv, h\to b\bar{b}$ at CEPC 240\,GeV are shown on the left panel of \autoref{fig:depm1}, after applying the selection cuts which include a Higgs-mass-window cut of $105\,{\rm GeV}<m^{\rm rec}_h<135\,{\rm GeV}$ on the $b$-jet pair.  
While $\delta_p$ has a slightly larger spread than $\delta_E$,  
its coefficients in \autoref{eq:dmz} are much smaller.  We therefore expect the distribution $\mrcp$ to have the smallest spread, and the one of $\mrce$ to have the largest.  This is verified on the right panel of \autoref{fig:depm1} where the distributions of $\mrc$, $\mrce$ and $\mrcp$ are shown.
For the $WW$-fusion events, we expect a less significant difference among the distributions of the three variables (which are shown later in \autoref{fig:nfitilc} \& \ref{fig:nfitcepc}), since they do not have a $Z$ in the event.
The corresponding distributions for ILC 250\,GeV are very similar to the ones in \autoref{fig:depm1}.

It should be noted that the distribution of $\delta_E$ in \autoref{fig:depm1} is asymmetric, suggesting that on average the measured energy of the $b$-jet pair is smaller than its actual value.  This is due to the fact that in our simulation we do not apply any jet energy corrections that are widely used in the LHC experiments~\cite{Khachatryan:2016kdb, Aaboud:2017jcu}.  As a result, the central values of the $\mrc$ and $\mrce$ distributions are also shifted to be larger than $m_Z$.   Assuming a jet energy correction mechanism will be implemented in the future lepton collider(s), one would expect a corrected central value and also some improvements on the energy measurement, and the $\mrc$ (and $\mrce$) distribution will have a peak value around $m_Z$ and a slightly smaller spread.  We do not expect the lack of jet energy correction to have a significant impact on our results since the $\mrcp$ distribution still has a much smaller spread due to the parametric suppression of it uncertainty near the $hZ$ threshold as discussed above.~\footnote{We thank Zhen Liu for very valuable discussions on the topic of jet energy corrections.}

Having found that the variable $\mrcp$ could better reconstruct the $Z$ mass than $\mrc$, we perform an analysis based on a fast simulation to 
explicitly exam their discriminating powers on the $WW$-fusion and $hZ$ events at the CEPC 240\,GeV (with unpolarized beams) and ILC 250\,GeV (assuming $P(e^-,e^+) = (-0.8,+0.3)$).  We generate events for both processes using \texttt{Madgraph5}~\cite{Alwall:2014hca}, which are showered with \texttt{Pythia}~\cite{Sjostrand:2006za} before passing to \texttt{Delphes}~\cite{deFavereau:2013fsa} with ILD cards (using the detector geometry and flavor tagging efficiencies given in Ref.~\cite{Behnke:2013lya}) for detector simulations.  The interference term between the $WW$-fusion and $hZ$ processes are ignored.  It should be noted that the effects of ISR photons are not considered in the simulation with \texttt{Madgraph5}.  However, we expect their effects to be much smaller compared with the ones of jet resolutions.  We use the ILC analysis in Ref.~\cite{Durig:2014lfa} as a guide to validate our results from the simple simulation.  While the Durham jet clustering algorithm is used in Ref.~\cite{Durig:2014lfa}, it pointed out that the anti-$k_t$ jet algorithm with jet radius $R=1.5$ has a similar performance in the Higgs invariant mass reconstruction, which is used in our simulation.   We also follow closely the selection cuts in Ref.~\cite{Durig:2014lfa}.  In particular, each event is required to have exactly two $b$-jets and a cut on the invariant mass of the $b$-jet pair, $105\,{\rm GeV}<m^{\rm rec}_h<135\,{\rm GeV}$, is applied to reduce the backgrounds.  
The cuts related to variables in the Durham jet clustering algorithm 
are replaced by the simple requirement on jet number ($=2$) in each event. 
After event selections, we scale the number of signal events of ILC 250\,GeV to the ones in Ref.~\cite{Durig:2014lfa} (normalized to $2\inab$).  A similar scaling is also applied for CEPC, taking count of the differences in cross sections and selection efficiencies between CEPC and ILC.

The composition of the background in the $\eevvh$ channel is also listed in Ref.~\cite{Durig:2014lfa}.  The major ones are $\nu\bar{\nu}b\bar{b}$ and $q\bar{q}$, which contributes to $42\%$ and $34\%$ of the total background after selection cuts.  The $q\bar{q}$ background is difficult to simulate due to its huge cross section and tiny selection efficiency.  For simplicity, we simulate only the $\nu\bar{\nu}b\bar{b}$ background, apply the selection cuts and scale it up to match the total background number, given in Ref.~\cite{Durig:2014lfa} and normalized to our run scenarios.  We expect this simple treatment to provide a reasonable estimation on the effects of the backgrounds.


After selection cuts, the $\mrc$ and $\mrcp$ distributions of $hZ$ ($Z\to \nu\bar{\nu}$), $WW$-fusion and background events  are shown in \autoref{fig:nfitilc} for ILC 250\,GeV ($2\inab$ data with $P(e^-,e^+)=(-0.8,+0.3)$) and \autoref{fig:nfitcepc} for CEPC 240\,GeV ($5\inab$ data, unpolarized beams).  
In \autoref{fig:nfitilc} and \autoref{fig:nfitcepc}, the $\mrc$ ($\mrcp$) distribution is shown on the left (right) panel, while the distributions in the bottom panels are simply the amplified versions of the ones in the top panel.  To check the validity of our results, we compared our $\mrc$ distributions to the ones in Refs.~\cite{Durig:2014lfa, CEPC-SPPCStudyGroup:2015csa} and found a reasonable agreement in terms of spread ranges and distribution shapes.  Comparing the $\mrc$ and $\mrcp$ distributions in \autoref{fig:nfitilc}~and~\ref{fig:nfitcepc}, it is clear that $\mrcp$ provides a better discrimination between $hZ$ and $WW$-fusion events, with the $\mrcp$ distribution of $hZ$ having a much sharper peak around the $Z$ mass than the $\mrc$ one.  On the other hand, the $\mrcp$ distribution of the $\nu\bar{\nu}b\bar{b}$ background has a more even spread than the $\mrc$ one.  This is because that for the $\nu\bar{\nu}b\bar{b}$ background the $b\bar{b}$ pair does not come from the Higgs decay.  Using the wrong mass assumption therefore makes the reconstruction of $Z$ mass worse.   

Following Ref.~\cite{Durig:2014lfa}, we apply a $\chi^2$ fit to the binned $\mrc$ and $\mrcp$ distributions to extract the precisions (one-sigma uncertainties) of the $hZ$ and $WW$ cross sections, denoted as $\sigma_{hZ}$ and $\sigma_{WW\to h}$.  This is done by treating the overall rates, $\sigma_{hZ}$ and $\sigma_{WW\to h}$ as free parameters while assuming a perfect knowledge on the the shapes of the distributions.  Ref.~\cite{Durig:2014lfa} also treats the overall cross section of the background ($\sigma_{\rm bg}$) as a free parameter.  We consider two cases with $\sigma_{\rm bg}$ treated as a free parameter as well as fixing it to the predicted value.  The total $\chi^2$ of the $\mrc$ or $\mrcp$ distribution is given by
\begin{equation}
\chi^2 = \sum_i \, \frac{(n^i_{\rm theory} - n^i_{\rm exp})^2}{n^i_{\rm exp} } \,, 
\end{equation}
where for each bin $i$, $n^i_{\rm exp}$ is the expected number of events from simulation and $n^i_{\rm theory}$ is a function of $\sigma_{hZ}$ and $\sigma_{WW\to h}$ (and $\sigma_{\rm bg}$).  To ensure enough statistics in each bin, we choose a bin size of $5\,$GeV except for the first and last bin, which are chosen to include all the events below 65\,GeV (60\,GeV) and above 130\,GeV (115\,GeV) for the $\mrc$ ($\mrcp$) distribution at ILC, and all the events below 65\,GeV (60\,GeV) and above 120\,GeV (110\,GeV) for the $\mrc$ ($\mrcp$) distribution at CEPC.  The results of the $\chi^2$ fits are presented in \autoref{tab:nfitmmilc} for ILC 250\,GeV ($2\inab$ data with $P(e^-,e^+)=(-0.8,+0.3)$) and \autoref{tab:nfitmmcepc} for CEPC 240\,GeV ($5\inab$ data, unpolarized beams).

Overall, our results on the precision of the $WW$-fusion cross section is slightly worse than the ones in Ref.~\cite{Durig:2014lfa} (if normalized to the same luminosity) and the CEPC preCDR~\cite{CEPC-SPPCStudyGroup:2015csa}.  This is not surprising, since the results could depend on details of the simulation, for which we only performed a simplified study.  In what follows, we shall focus on the relative difference between the results from $\mrc$ and $\mrcp$.  While the distributions of $\mrcp$ clearly better separates the $hZ$ and $WW$-fusion events as shown in \autoref{fig:nfitilc} and \autoref{fig:nfitcepc}, in the 3-parameter fit (shown on the left panels of \autoref{tab:nfitmmilc} and \autoref{tab:nfitmmcepc}) the precision of $\sigma_{WW\to h}$ from the $\mrcp$ distribution is similar to (or even worse than) the one of $\mrc$, due to the fact that $WW$-fusion and background events have a larger overlap in the $\mrcp$ distribution than in the $\mrc$ one.  $\mrcp$ nevertheless significantly improves the precisions of $\sigma_{hZ}$ and $\sigma_{\rm bg}$ in the 3-parameter fit.  Assuming a good knowledge of the background, one may also fix the background cross section to the predicted value.  In the 2-parameter fit with $\sigma_{hZ}$ and $\sigma_{WW\to h}$, the $\mrcp$ distribution indeed provide a significantly better constraint on $\sigma_{WW\to h}$, with an improvement  of about $30\%$ at ILC and $20\%$ at CEPC, compared with the constraint from the $\mrc$ distribution.
\begin{table}[t]
\small
\centering
\begin{tabular}{|c||c|ccc||c|cc|} \hline
\multicolumn{8}{|c|}{ILC 250\,GeV, $2\inab$, uncertainties normalized to SM predictions}  \\  \hline \hline
& \multicolumn{4}{c||}{ 3-parameter fit } &  \multicolumn{3}{c|}{ fixing $\sigma_{\rm bg}$ } \\  \cline{2-8}
$\mrc$ & uncertainty &   \multicolumn{3}{c||}{correlation matrix}  &  uncertainty  &   \multicolumn{2}{c|}{correlation matrix}   \\ \cline{2-8}
&  &    $\sigma_{hZ}$  & $\sigma_{WW\to h}$  & $\sigma_{\rm bg}$  &  &     $\sigma_{hZ}$  & $\sigma_{WW\to h}$     \\ \hline
$\sigma_{hZ}$                       & 0.049  &    1 & 0.47 & -0.97      &    0.011          &    1 & -0.69      \\
$\sigma_{WW\to h}$              & 0.063  &      & 1 & -0.63        &    0.045          &       &  1            \\
$\sigma_{\rm bg}$                 & 0.31 &         &   & 1               &                  &       &                    \\  \hline \hline
 & \multicolumn{4}{c||}{ 3-parameter fit } &  \multicolumn{3}{c|}{ fixing $\sigma_{\rm bg}$ } \\  \cline{2-8}
$\mrcp$  & uncertainty &   \multicolumn{3}{c||}{correlation matrix}  &  uncertainty  &   \multicolumn{2}{c|}{correlation matrix}   \\ \cline{2-8}
 &  &    $\sigma_{hZ}$  & $\sigma_{WW\to h}$  & $\sigma_{\rm bg}$  &  &     $\sigma_{hZ}$  & $\sigma_{WW\to h}$     \\ \hline
$\sigma_{hZ}$                       & 0.010  &     1 & 0.21& -0.51     &    0.0088          &    1 & -0.46      \\
$\sigma_{WW\to h}$              & 0.059  &       & 1 & -0.83            &    0.033          &       &  1            \\
$\sigma_{\rm bg}$                 & 0.088  &         &      &  1               &                   &       &                    \\  \hline
\end{tabular}
\caption{The one sigma uncertainties and correlations of the cross sections of the Higgsstrahlung process with an invisible $Z$ ($\sigma_{hZ}$) and the $WW$-fusion process ($\sigma_{WW\to h}$) at ILC 250\,GeV from a fit to the $\mrc$ (top panel) and $\mrcp$ (bottom panel) distributions.  A total luminosity of $2\inab$ with beam polarization of $P(e^-,e^+) = (-0.8,+0.3)$ is assumed.  
In the 3-parameter fit on the left panel, the overall normalization of the background is treated as a free parameter in the fit.  In the 2-parameter fit on the left panel, the total number of background events is fixed to the predicted value.}
\label{tab:nfitmmilc}
\end{table}
\begin{table}[t]
\small
\centering
\begin{tabular}{|c||c|ccc||c|cc|} \hline
\multicolumn{8}{|c|}{CEPC 240\,GeV, $5\inab$, uncertainties normalized to SM predictions}  \\  \hline \hline
& \multicolumn{4}{c||}{ 3-parameter fit } &  \multicolumn{3}{c|}{ fixing $\sigma_{\rm bg}$ } \\  \cline{2-8}
$\mrc$ & uncertainty &   \multicolumn{3}{c||}{correlation matrix}  &  uncertainty  &   \multicolumn{2}{c|}{correlation matrix}   \\ \cline{2-8}
&  &    $\sigma_{hZ}$  & $\sigma_{WW\to h}$  & $\sigma_{\rm bg}$  &  &     $\sigma_{hZ}$  & $\sigma_{WW\to h}$     \\ \hline
$\sigma_{hZ}$                       & 0.024  &   1 & 0.28& -0.95      &    0.0077          &    1 & -0.61      \\
$\sigma_{WW\to h}$              & 0.058  &      & 1 & -0.47      &    0.051          &       &  1            \\
$\sigma_{\rm bg}$                 & 0.20 &     &           &  1           &                  &       &                    \\  \hline \hline
 & \multicolumn{4}{c||}{ 3-parameter fit } &  \multicolumn{3}{c|}{ fixing $\sigma_{\rm bg}$ } \\  \cline{2-8}
$\mrcp$  & uncertainty &   \multicolumn{3}{c||}{correlation matrix}  &  uncertainty  &   \multicolumn{2}{c|}{correlation matrix}   \\ \cline{2-8}
 &  &    $\sigma_{hZ}$  & $\sigma_{WW\to h}$  & $\sigma_{\rm bg}$  &  &     $\sigma_{hZ}$  & $\sigma_{WW\to h}$     \\ \hline
$\sigma_{hZ}$                       & 0.0071  &    1 & 0.098 & -0.35     &    0.0066          &    1 & -0.45      \\
$\sigma_{WW\to h}$              & 0.083 &        & 1 & -0.87             &    0.041          &       &  1            \\
$\sigma_{\rm bg}$                 & 0.082  &         &      &  1               &                  &       &                    \\  \hline
\end{tabular}
\caption{Same as \autoref{tab:nfitmmilc} but for CEPC 240\,GeV with $5\inab$ data and unpolarized beams.}
\label{tab:nfitmmcepc}
\end{table}

In the \emph{kappa} framework with the $hZZ$ and $hWW$ couplings treated as independent parameters, the $WW$-fusion measurement is an important input for constraining the $hWW$ coupling (with the other being the Higgs decay, $h\to WW^*$).  With a 20-30\% improvement on the precision of the $WW$-fusion cross section,  a sizable improvement for the constraint on the $hWW$ coupling is expected.  The $WW$-fusion cross section is also an important input for the determination of the Higgs total width, following the relation~\cite{CEPC-SPPCStudyGroup:2015csa}\footnote{It should be noted that, despite the usual claim of being model independent, \autoref{eq:width} explicitly assumes that the $hWW$ coupling is independent of the energy scale ({\it i.e.}, anomalous couplings such as $hW^{\mu\nu}W_{\mu\nu}$ are absent), which is not true under the more general EFT framework.}
\begin{equation}
\Gamma_h \propto \frac{\Gamma(h\to b\bar{b})}{{\rm BR}(h\to b\bar{b})}  \propto \frac{\sigma(\vvh, \,h\to b\bar{b})}{{\rm BR}(h\to b\bar{b}) \cdot {\rm BR}(h\to WW^*)} \,,  \label{eq:width}
\end{equation}
where, following the usual convention, $\sigma(\vvh)$ denotes only the $WW$-fusion contribution to it. With a 20-30\% improvement on the precision of $\sigma(\vvh, \,h\to b\bar{b})$, as well as some possible improvement on the determination of ${\rm BR}(h\to b\bar{b})$ from a better measurement of $\sigma(hZ, \,Z\to \nu \bar{\nu}, \,h \to b\bar{b})$, the precision of the Higgs total width obtained using \autoref{eq:width} could be improved by at least 20-30\% using the $\mrcp$ variable.

To conclude this section, we would like to emphasize that, while we try our best to validate our results, they do rely on simple simulations and should be explicitly tested by experimental groups with proper simulation tools.  The fits performed in obtaining the results in \autoref{tab:nfitmmilc} and \ref{tab:nfitmmcepc} also assume a perfect knowledge of the distribution shapes for each process, which may not be a good assumption in an actual experiment.  We also include only the $\nu\bar{\nu}b\bar{b}$ background, while other backgrounds may have different kinematic features.  Nevertheless, we expect our results to still hold qualitatively due to the simple reasoning that $\mrcp$ has a smaller uncertainty and better reconstruct the $Z$ mass.


\section{Improving Higgs coupling constraints in the EFT framework}
\label{sec:heft}

Having explored the capability of the $\mrcp$ variable in improving the measurement of the $WW$-fusion process, we are now ready to exam its impact on the determination of the Higgs couplings.
We choose to study the Higgs coupling constraints in a global effective-field-theory (EFT) framework with dimension six (D6) operators.\footnote{For recent Higgs EFT studies in the contexts of future lepton colliders, see Refs.~\cite{Ellis:2015sca, Durieux:2017rsg, Barklow:2017suo, Craig:2014una, Beneke:2014sba, Henning:2014gca, Craig:2015wwr, Ellis:2017kfi, Barklow:2017awn}.}  Such a framework has several advantages.  First, assuming the scale of new physics is high, the EFT with D6 operators gives a good parameterization of the effects of new physics and the results can be mapped to any specific model that satisfies the assumptions of the framework.  Second, it takes count of the connections among different measurements.  For instance, some operators contribute to both Higgs processes and the dibson process, and the triple gauge coupling (TGC) measurements from the diboson process can thus help the overall constraints on the Higgs couplings~\cite{Falkowski:2015jaa}.  Gauge invariance is also imposed by construction in the EFT framework.  We focus on the CEPC 240\,GeV and follow  Ref.~\cite{Durieux:2017rsg} in terms of the basis choice and measurement inputs.  In particular, focusing on the Higgs and diboson measurements at 240\,GeV, and making reasonable assumptions, a total number of 11 parameters are sufficient to describe the contributions from the D6 operators.    A short summary of the framework in Ref.~\cite{Durieux:2017rsg} is provided in \autoref{app:eft}.  The methods we propose should nevertheless be applicable to other collider scenarios and frameworks.

A few important differences between the cross-section fit in \autoref{sec:mrecoil} and the EFT analysis should be noted.  While not specifically mentioned, the cross-section fit does make assumptions on the new physics, in particular that it only modifies the overall rates, not the differential distributions, of the $hZ$ and $WW$-fusion processes.  In the cross-section fit, the $hWW$ and $hZZ$ couplings are also assumed to be independent, regardless of their relation from gauge invariance.  The EFT analysis, while imposing gauge invariance, contain anomalous couplings of the form $h Z_{\mu\nu} Z^{\mu\nu}$ and $h Z_\mu \partial_\nu Z^{\mu\nu}$ (and the same for $W$) which have different momentum dependences from the SM couplings due to the extra derivatives.  The potential new physics contribution to the $hZ\gamma$ vertex is also included in the EFT analysis, which could contribute to $\eehz$ via an $s$-channel photon.  It is thus beter to directly fit the EFT parameters to the $\mrc$ or $\mrcp$ distribution of the inclusive $\eevvh$ process instead of fitting them to the extracted precisions of cross sections from \autoref{sec:mrecoil}.  By fitting to the inclusive $\eevvh$ process we also include the interference term of $WW$-fusion and $hZ$ processes, which is usually ignored in cross-section fits.  The expressions for the total cross section of $\eevvh$ and the (binned) differential ones of the $\mrc$ and $\mrcp$ distributions in terms of the EFT parameters are listed in \autoref{app:expression}.

The measurement inputs of the Higgsstrahlung ($\eehz$) and diboson ($\eeww$) processes at CEPC 240\,GeV are listed in \autoref{tab:cepcinput}.  The estimations of $hZ$ measurements are taken from Ref.~\cite{CEPCupdate2}, which updates the ones in the CEPC preCDR~\cite{CEPC-SPPCStudyGroup:2015csa}.  
In addition, The angular observables of $\eehz$ in Ref.~\cite{Beneke:2014sba} are included, for which we use only the channel $\eehz, \,h\to b\bar{b}, \,Z\to \ell^+\ell^-$ and assume a fixed 60\% signal selection efficiency, following Refs.~\cite{Craig:2015wwr, Durieux:2017rsg}.
For the TGC measurements, we follow the treatment in Ref.~\cite{Durieux:2017rsg} which adopts the one in Ref.~\cite{Bian:2015zha} with the addition of a universal $1\%$ systematical uncertainty in each bin of all differential distributions.  We directly list the resultant one-sigma constraints of the anomalous TGC parameters and their correlations in \autoref{tab:cepcinput}.  We construct the total $\chi^2$ by summing over the $\chi^2$s of all measurements and perform global fits to obtain the precision reaches (one-sigma bounds) of the relevant EFT parameters.

\begin{table}[t]\small
\centering
\begin{tabular}{|c|c||c|c|ccc|} \hline
  \multicolumn{7}{|c|}{CEPC 240\,GeV, $5\inab$, unpolarized beams}    \\  \hline 
 \multicolumn{2}{|c||}{$\eehz$} &  \multicolumn{5}{|c|}{$\eeww$}     \\  \hline    
$\sigma(\eehz)$  &  0.50\%  &   &   uncertainty &   \multicolumn{3}{c|}{correlation matrix}        \\  \hline\hline
&  $\sigma(hZ) \times {\rm BR}$  &              &  &    $\delta g_{1,Z}$  &  $\delta \kappa_\gamma$  & $\lambda_Z$      \\  \hline
$h\to b\bar{b}$  & $0.24\%^\bigstar$    &    $\delta g_{1,Z}$                & $6.4\times 10^{-3}$  &  1 & 0.068 & -0.93         \\ 
$h\to c\bar{c}$  & 2.5\%    &  $\delta \kappa_\gamma$  & $3.5\times 10^{-3}$  &     & 1        & -0.40     \\ 
$h\to gg$          & 1.2\%     &   $\lambda_Z$                     & $6.3\times 10^{-3}$  &     &           &  1          \\  \cline{3-7}
$h\to \tau\tau$  & 1.0\%    &     \multicolumn{5}{c|}{}     \\ 
$h\to WW^*$    & 1.0\%    &     \multicolumn{5}{c|}{  }    \\ 
$h\to ZZ^*$      & 4.3\%     &     \multicolumn{5}{l|}{Angular observables in }     \\ 
$h\to \gamma\gamma$ & 9.0\%    &     \multicolumn{5}{l|}{$\eehz, \,h\to b\bar{b}, \,Z\to \ell^+\ell^-$}     \\ 
$h\to \mu\mu$  & 12\%        &    \multicolumn{5}{l|}{are also included.}     \\ 
$h\to Z\gamma$ & 25\%     &     \multicolumn{5}{c|}{}      \\   \hline
\end{tabular}
\caption{A summary of the measurement inputs from $\eehz$ and $\eeww$ at CEPC 240\,GeV used in the EFT fit, assuming a total luminosity of $5\inab$ and unpolarized beams.  Inputs on rate measurements of $\eehz$ are from Ref.~\cite{CEPCupdate2} which updates the estimations in the preCDR~\cite{CEPC-SPPCStudyGroup:2015csa}.   For the precision of $\sigma(hZ) \times{\rm BR}(h\to b\bar{b})$ (marked by a star $^\bigstar$), we have excluded the contribution from $\eehz, Z\to\nu\bar{\nu}, h\to b\bar{b}$ to avoid double counting with $\eevvh, h\to b\bar{b}$.  The angular observables in $\eehz, \,h\to b\bar{b}, \,Z\to \ell^+\ell^-$ are included, assuming a $60\%$ signal selection efficiency.  The constraints on aTGC parameters from measurements of $\eeww$ are obtained following the treatments in Ref.~\cite{Durieux:2017rsg}.
}  
\label{tab:cepcinput}
\end{table}

\begin{figure}[t]
\centering
\includegraphics[width=0.9\textwidth]{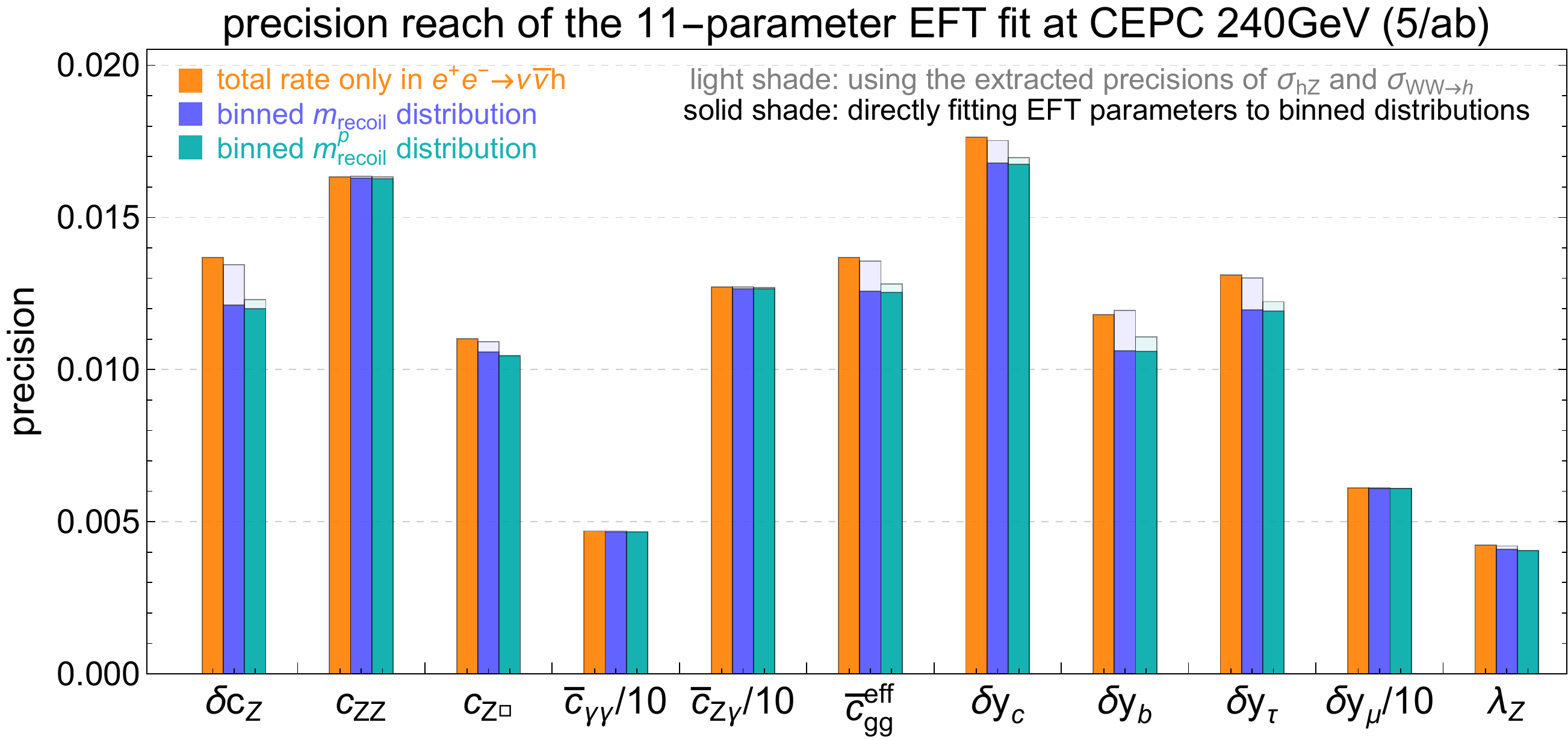}
\caption{The one-sigma precision reach of the 11-parameter fit in the EFT framework at CEPC 240\,GeV with $5\inab$ data and unpolarized beams.  See \autoref{app:eft} for the definitions of parameters.  Three scenarios are shown which differ on the information used for the $\eevvh$ measurement.  The first one uses only the total rate of $\eevvh$.  The second (third) one uses the $\mrc$ ($\mrcp$) distribution, and contain two sub-scenarios.  The one shown by the light shaded columns uses the extracted precisions of $\sigma_{hZ}$ and $\sigma_{WW\to h}$ in \autoref{sec:mrecoil} (the two-parameter fit on the right panel of \autoref{tab:nfitmmcepc}, correlation ignored).  The one shown by the solid columns is obtained from a direct fit to the binned $\mrc$ ($\mrcp$) distribution.  }
\label{fig:barcepc1}
\end{figure}

We consider three scenarios in the global analysis.  All three use the inputs on Higgsstrahlung and TGC measurements in \autoref{tab:cepcinput}, but different information on the measurement of $\eevvh$.  The first one uses only the total rate of $\eevvh$.  The second (third) uses the information in the $\mrc$  ($\mrcp$) distribution, with the EFT parameters directly fitted to the binned distributions.  We also compare the reach with the more conventional method of fitting the EFT parameters to the extracted precisions of $hZ$ and $WW$-fusion cross sections in \autoref{sec:mrecoil}, ignoring the correlation between the two cross sections (which is often not reported).  
Since the background is assumed to be SM-like in the EFT analysis,\footnote{It is reasonable to fix the background (which has no Higgs) to the SM predictions in an EFT global framework in this case, as deviations from SM are strongly constrained by the electroweak precision measurements at $Z$-pole or other measurements.  Fixing the background nevertheless requires one to have a very good knowledge of the total rate and distribution shape of the background as pointed out in \autoref{sec:mrecoil}. } for the extracted precision of cross sections we also use the results of the two-parameter fit with fixed background, shown on the right panel of \autoref{tab:nfitmmcepc}.  The results of the 11-parameter fit are presented in \autoref{fig:barcepc1}.

Comparing the reaches of the three scenarios (shown by the orange, blue and cyan columns in \autoref{fig:barcepc1}), we indeed observe a non-negligible improvement on the overall reach if the information in the $\mrc$ or $\mrcp$ distribution is used.  In particular, the reach on the parameter $\delta c_Z$ (corresponding to a shift in the SM $hZZ$ and $hWW$ couplings) is improved by more than 10\%.  Consequently, the reach on $\bar{c}^{\rm \,eff}_{gg}$, $\delta y_c$, $\delta y_b$ and $\delta y_\tau$, which contribute to the Higgs decay to $gg$, $c\bar{c}$, $b\bar{b}$ and $\tau\bar{\tau}$, have also been improved by a similar (or slightly less) factor.  The reach with the $\mrcp$ distribution is better than the one with $\mrc$ as we expected.  However, the relative improvement from $\mrc$ to $\mrcp$ turns out to be very marginal.  We also find that, if fitting the EFT parameters to the extracted cross sections $\sigma_{hZ}$ and $\sigma_{WW\to h}$ without taking count of their correlation (the results are shown with light shades for the 2nd and 3rd columns), the reaches are worse than the ones from direct fitting the EFT parameters to the distributions, in particular for the $\mrc$ distribution.  This is because, as shown in \autoref{sec:mrecoil}, the uncertainties of $\sigma_{hZ}$ and $\sigma_{WW\to h}$ have a large correlation between them due to the difficulty in separating the two, in particular for the $\mrc$ distrubiton.  This correlation is usually not reported in official documents, and the omission of it could lead to a considerable impact on the overall reach.
It should be noted that, our results using only the total rate of $\eevvh$ is worse than the corresponding ones in Ref.~\cite{Durieux:2017rsg}.  This is because our estimations on the rate measurement of $\eevvh$ is more conservative than the one in Ref.~\cite{Durieux:2017rsg}, which is derived from the CEPC preCDR~\cite{CEPC-SPPCStudyGroup:2015csa}.    
If the overall cross section measurement of $\eevvh$ can be improved ({\it e.g.} by optimizing the selection cuts), we expect the use of $\mrc$ and $\mrcp$ distributions would also bring a more significant improvement on the overall reach of the EFT fit.  We have also chosen very conservative bin sizes to control the uncertainties in each bin from simulation.  Further optimizations of the analysis may also provide substantial improvements on the reach with the $\mrc$ and $\mrcp$ distributions.

We also find that, if the TGCs can be measured with much better precisions such as the ones in Ref.~\cite{Barklow:2017suo} for ILC 250\,GeV (which are one order of magnitude better than the ones in \autoref{tab:cepcinput}), or if  
multiple runs with different beam polarizations are available (also likely to be the case for ILC), the improvement from the $\mrc$ and $\mrcp$ distributions with respect to using only the total rate of $\eevvh$ becomes rather insignificant.  This is not surprising, since very precise TGC measurements can effectively remove two degrees of freedom in the fit so that there is less need for additional handles to discriminate the parameters.  The interference term of the $\eehz$ diagram with an $s$-channel $Z$ and the one with an $s$-channel photon is also sensitive to the beam polarization, which can help probe the operators that contribute to this interference~\cite{Durieux:2017rsg, Barklow:2017suo}.  In general, once a sufficient number of constraints are included in a global analysis, the overall precision reach is expected to be less sensitive to the impact of a single measurement, such as the one of $\eevvh$. 
It is nevertheless important to optimize the measurements in order to maximize the sensitivity to new physics.


\section{Applications on the inclusive $hZ$ measurements}
\label{sec:hz}

While we have focused on the $WW$-fusion measurements in the previous two sections, a question of great interest is whether the variable $\mrcp$ could be applied to improve the inclusive measurement of $\eehz$, where the decay product of $Z$ are tagged instead.  
While suffering from the jet resolution, the hadronic $Z$ decay channel provides a slightly better measurement on $\sigma(Zh)$ than the leptonic one, thanks to its large branching ratio.\footnote{For instance, the precision of the inclusive $hZ$ cross section measured from the leptonic (hadronic) $Z$ channel is reported to be 0.8\% (0.65\%) in the CEPC preCDR~\cite{CEPC-SPPCStudyGroup:2015csa}. }  An improved measurement of the hadronic $Z$ channel could thus have a significant impact on the overall precision reach of $\sigma(hZ)$.
It is straightforward to write down the recoil mass and its two variations for the reconstruction of the Higgs mass in $\eehz$, which are
\begin{align}
\mrc^2 =&~ s-2 \sqrt{s}\, E^{\rm rec}_Z + (m^{\rm rec}_Z)^2 \,,  \nonumber\\
(\mrce)^2 =&~ s-2 \sqrt{s}\, E^{\rm rec}_Z + m^2_Z \,,    \nonumber\\
(\mrcp)^2 =&~ s-2 \sqrt{s}\, \sqrt{m^2_Z+|\vec{p}_Z^{\rm \, rec}|^2} + m^2_Z \,,
\end{align}
where $E^{\rm rec}_Z$, $\vec{p}_Z^{\rm \, rec}$ and $m^{\rm rec}_Z$ are the reconstructed energy, 3-momentum and invariant mass of the $Z$ from the two jets, while $m_Z$ is the true $Z$ mass, fixed to be 91.19\,GeV.  Similar to \autoref{eq:dmz}, we derive the deviations in the measured $\mrc$, $\mrce$, $\mrcp$ as a function of the deviations in the measured $Z$ energy and 3-momentum to be
\begin{equation}
\delta_m ~/~ \delta_m^E ~/~ \delta_m^p \approx  
\left\{ \begin{matrix} 
-0.91 \, \delta_E -0.17 \, \delta_p ~/~ -1.6 \, \delta_E ~/~  -0.39 \, \delta_p \hspace{1cm}  \mbox{ at } 240\,{\rm GeV} \\  
-0.99 \, \delta_E -0.25 \, \delta_p ~/~  -1.8\, \delta_E ~/~  -0.56 \, \delta_p  \hspace{1cm}  \mbox{ at } 250\,{\rm GeV} 
\end{matrix} \right. \,,  \label{eq:dmh}
\end{equation}
where  $\delta_m$, $\delta_m^E$ and $\delta_m^p$ are defined as 
\begin{equation}
\mrc =~ \mrc^{\rm true} (1+\delta_m) \,,~~~~  \mrce = \mrc^{\rm true} (1+\delta_m^E) \,,~~~~   \mrcp = \mrc^{\rm true} (1+\delta_m^p) \,,  \label{eq:dmmm}
\end{equation}
with $\mrc^{\rm true} = m_h$.  For $\delta_E$ and $\delta_p$, the definitions are
\begin{equation}
E^{\rm rec}_Z  =~ E_Z (1+\delta_E) \,,~~~~~~~~~~~   |\vec{p}_Z^{\rm \, rec}| = |\vec{p}_Z| (1+\delta_p) \,,  \label{eq:depz}
\end{equation}
\begin{figure}[t]
\centering
\includegraphics[width=0.46\textwidth]{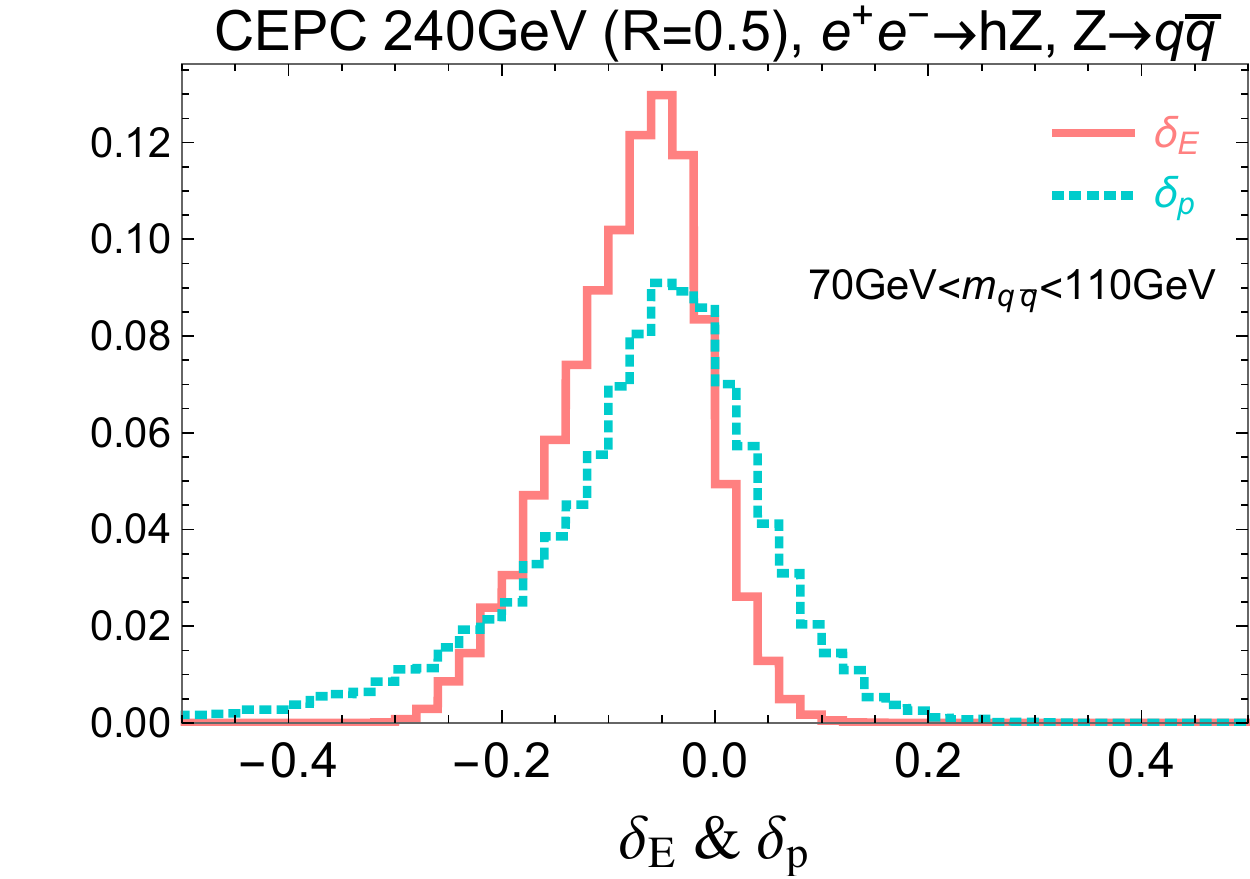} \hspace{0.1cm}
\includegraphics[width=0.46\textwidth]{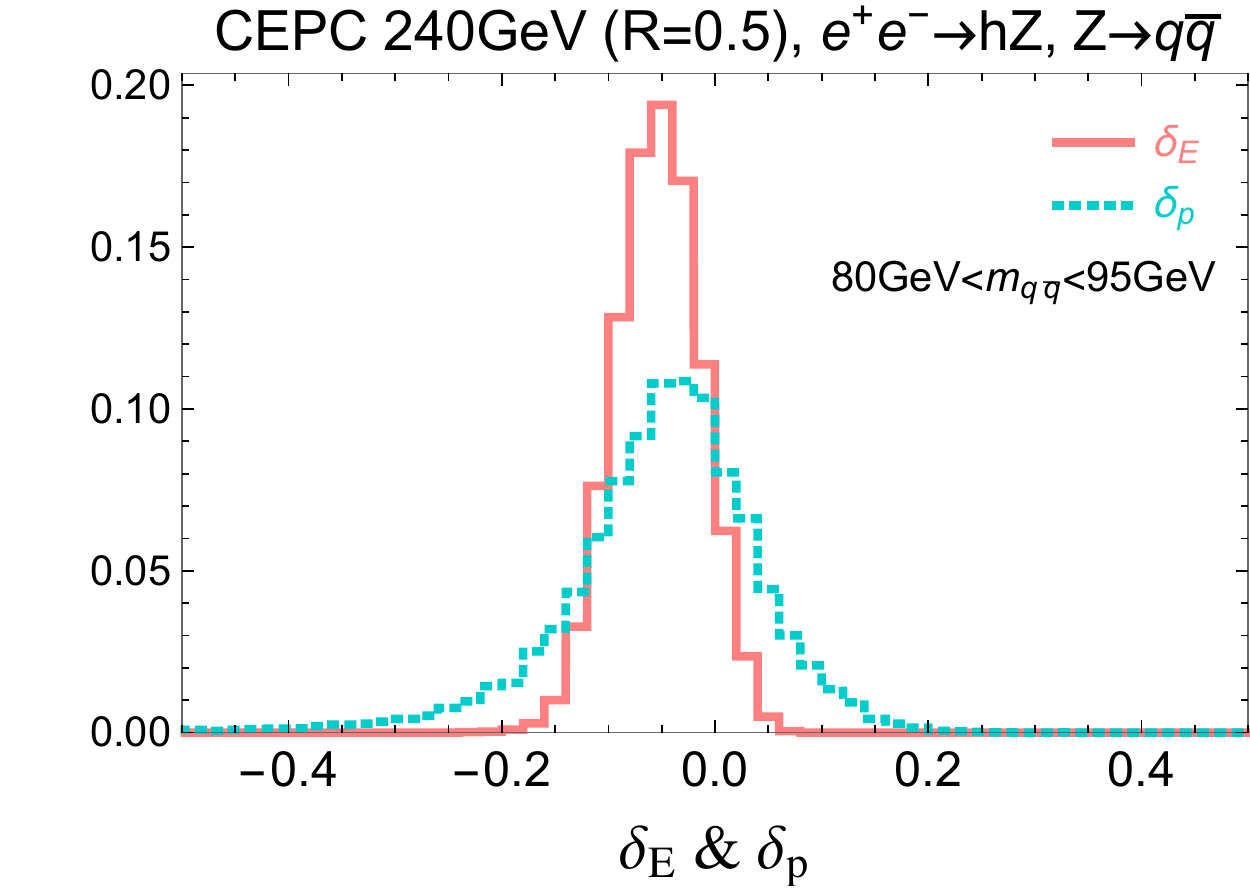}
\caption{The distributions of $\delta_E$ and $\delta_p$ of the reconstructed $Z$ (defined in \autoref{eq:depz}) in $\eehz, Z\to q\bar{q}$ at CEPC 240\,GeV after applying a $Z$-mass-window cut of $70\,{\rm GeV}<m_{q\bar{q}}<110\,{\rm GeV}$ (left) or $80\,{\rm GeV}<m_{q\bar{q}}<95\,{\rm GeV}$ (right).  The Higgs is forced to decay invisibly in the simulation to ensure the correct reconstruction of $Z$.  A radius of $R=0.5$ is used in the jet clustering algorithm. }
\label{fig:depmh1}
\end{figure}
where $E_Z$ and $\vec{p}_Z$ are the true energy and 3-momentum of the $Z$.  Similar to \autoref{eq:dmz}, in \autoref{eq:dmh} the coefficients of $\delta_p$ is also smaller than the ones of $\delta_E$, but with a suppression factor of $\sim |\vec{p}_Z|^2/E^2_Z$ instead.   
The distributions of $\delta_E$ and $\delta_p$ for the reconstructed $Z$ are shown in \autoref{fig:depmh1}, with the details of simulation stated later in this section.  Note that the cut on the $Z$-mass window has a strong impact on the distributions of $\delta_E$ and $\delta_p$.  For larger deviations of the measured energy and momentum from the true ones, the invariant mass also tends to be further away from its true value.  

\begin{figure}[t]
\centering
\includegraphics[width=0.47\textwidth]{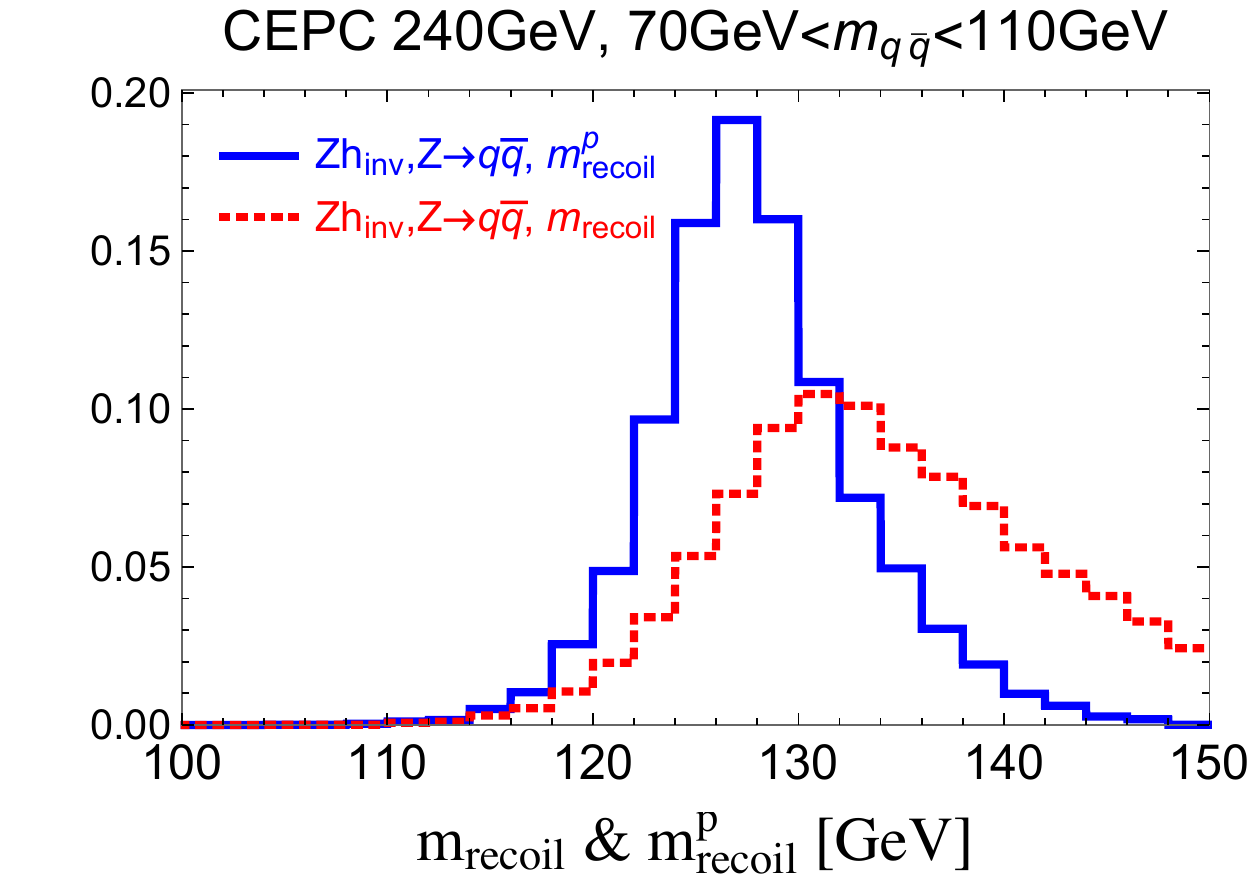} \hspace{0.1cm}
\includegraphics[width=0.47\textwidth]{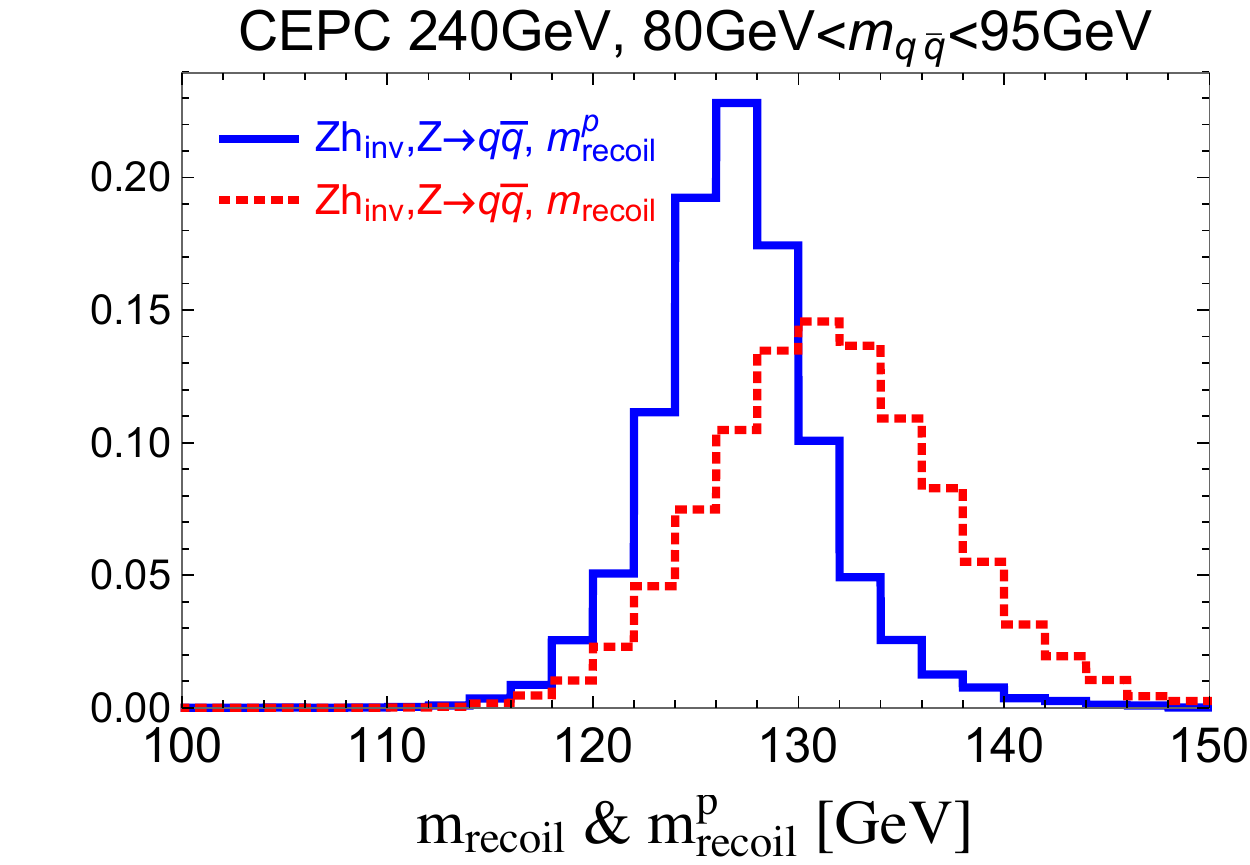}  \\ \vspace{0.3cm}
\includegraphics[width=0.47\textwidth]{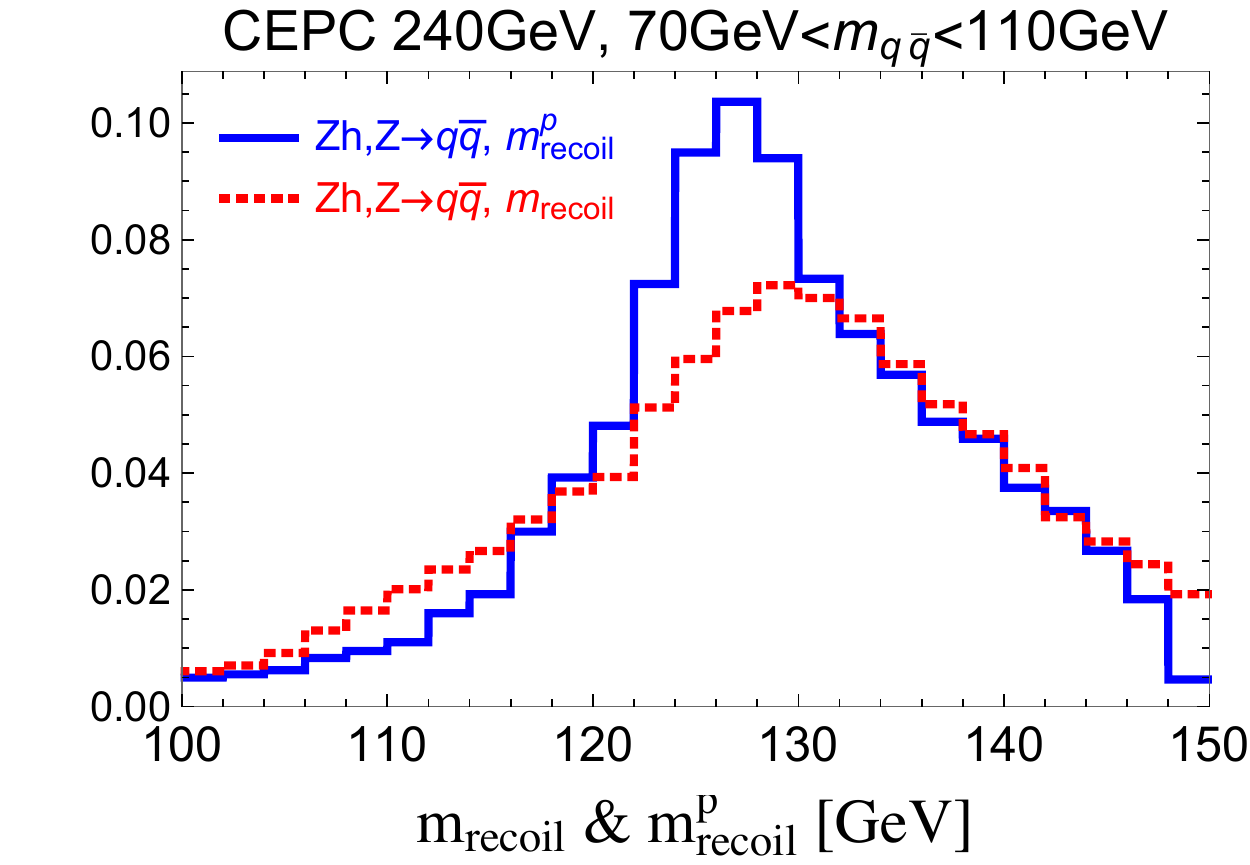} \hspace{0.1cm}
\includegraphics[width=0.47\textwidth]{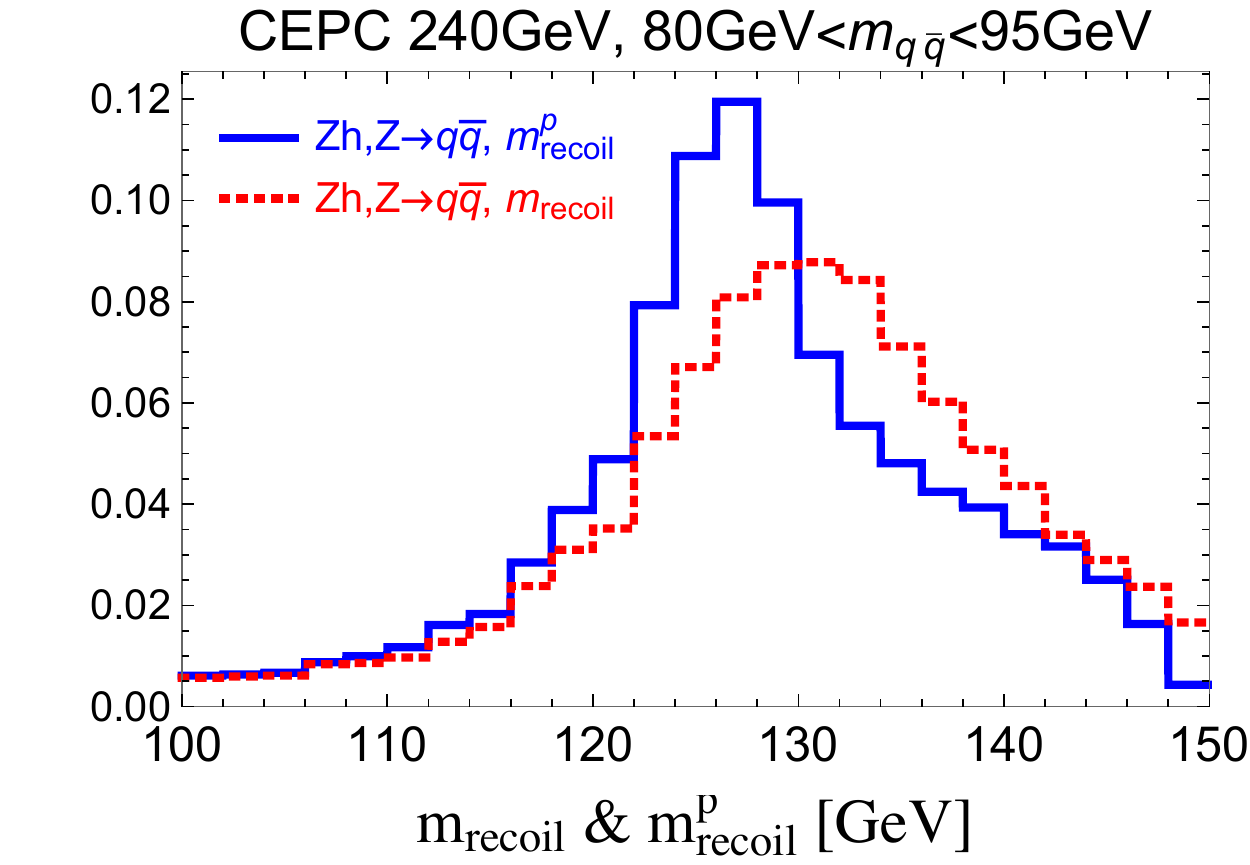}  
\caption{{\bf Top row:} The distributions of  $\mrc$ and $\mrcp$ for $\eehz, Z\to q\bar{q}$ at CEPC 240\,GeV.  The Higgs is forced to decay invisibly in the simulation to avoid the combinatorial problem.  {\bf Bottom row:} The same distributions with Higgs inclusive decay.  The left and right panels differ on the $Z$-mass-window cut, which is $70\,{\rm GeV}<m_{q\bar{q}}<110\,{\rm GeV}$ ($80\,{\rm GeV}<m_{q\bar{q}}<95\,{\rm GeV}$) for the left (right) panels.}
\label{fig:zhlw}
\end{figure}

To compare the reconstruction power of $\mrc$ and $\mrcp$ on the Higgs mass, we perform a simple analysis using the simulation tools listed in \autoref{sec:mrecoil}.  One important difference here is that for Higgs inclusive measurement with $Z\rightarrow q\bar{q}$, the final states could contain additional jets from Higgs decay, making it more difficult to reconstruct the $Z$.  Due to the additional jets, we set the jet radius to $R=0.5$ in order to reduce the contamination among the jets.  For an event with more than two jets, we choose the pair of jets with an invariant mass that is closest to the value of $Z$ mass.  We then apply a $Z$-mass-window cut on the invariant mass of the jet pair, $m_{q\bar{q}}$, intended for removing backgrounds.  The difference between $\mrc$ and $\mrcp$ is strongly correlated with the size of the $Z$ window -- in the limit that the invariant mass equals the actual $Z$ mass, $\mrc$ and $\mrcp$ become equivalent.  We therefore consider both a larger window, $70\,{\rm GeV}<m_{q\bar{q}}<110\,{\rm GeV}$, and a smaller one, $80\,{\rm GeV}<m_{q\bar{q}}<95\,{\rm GeV}$.    
The distributions of $\mrc$ and $\mrcp$ for $\eehz, Z\to q\bar{q}$ after the selection cuts are shown in \autoref{fig:zhlw} for CEPC 240\,GeV.  To estimate  the impact of the combinatorial problem in the reconstruction of $Z$, we first consider a case in which the Higgs is forced to decay invisibly in the simulation.  The only purpose of the invisible decay is to avoid having additional jets from the Higgs decay and ensure a clear identification of the $Z$ jet-pair.  The results are shown on the top panels of \autoref{fig:zhlw} for the two choices of $Z$-mass-window cuts.  For this ideal case, it is clearly that $\mrcp$ has a significantly narrower spread and provides a much better reconstruction of the Higgs mass than $\mrc$ does.  
The improvement with $\mrcp$ is more significant if a large $Z$-mass window cut is applied as we expected.  For the realistic case with Higgs inclusive decays, the distributions are shown in the bottom panels of \autoref{fig:zhlw}.  The reconstruction of the Higgs mass is worse for both $\mrc$ and $\mrcp$ distributions due to the wrong jet-pairing.  However, $\mrcp$ still has a better performance than $\mrc$, so its usefulness is not washed out by the combinatorial problem.  
We also note that, due to the lack of jet energy correction mentioned in \autoref{sec:mrecoil}, our distributions of $\mrc$ peak around $130$\,GeV rather than $125$\,GeV.  While the central values of the distributions can be corrected, we expect  $\mrcp$ to still have a better performance than $\mrc$ after the implementation of jet energy corrections due to the parametric suppression on the uncertainties of $\mrcp$ near the $hZ$ threshold.  
\begin{figure}[t]
\centering
\includegraphics[width=0.47\textwidth]{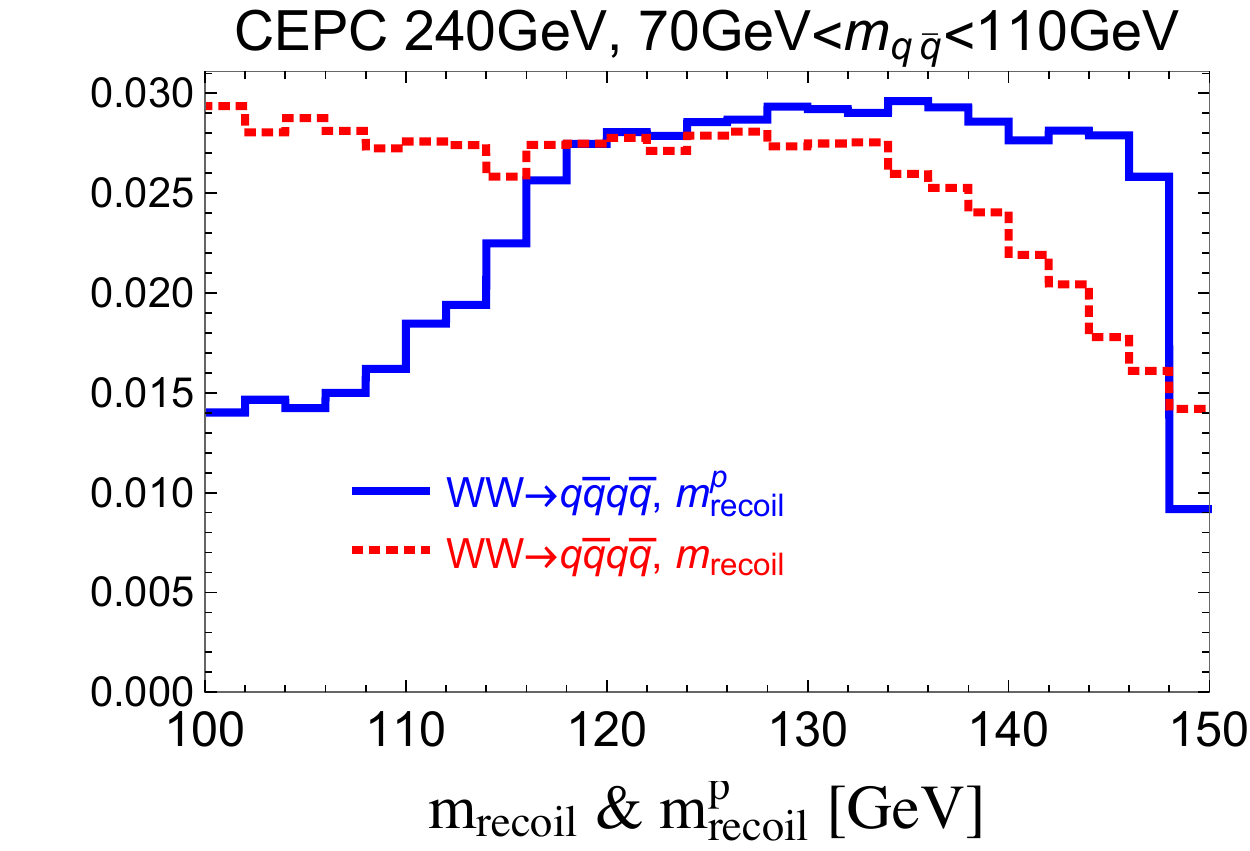} \hspace{0.1cm}
\includegraphics[width=0.47\textwidth]{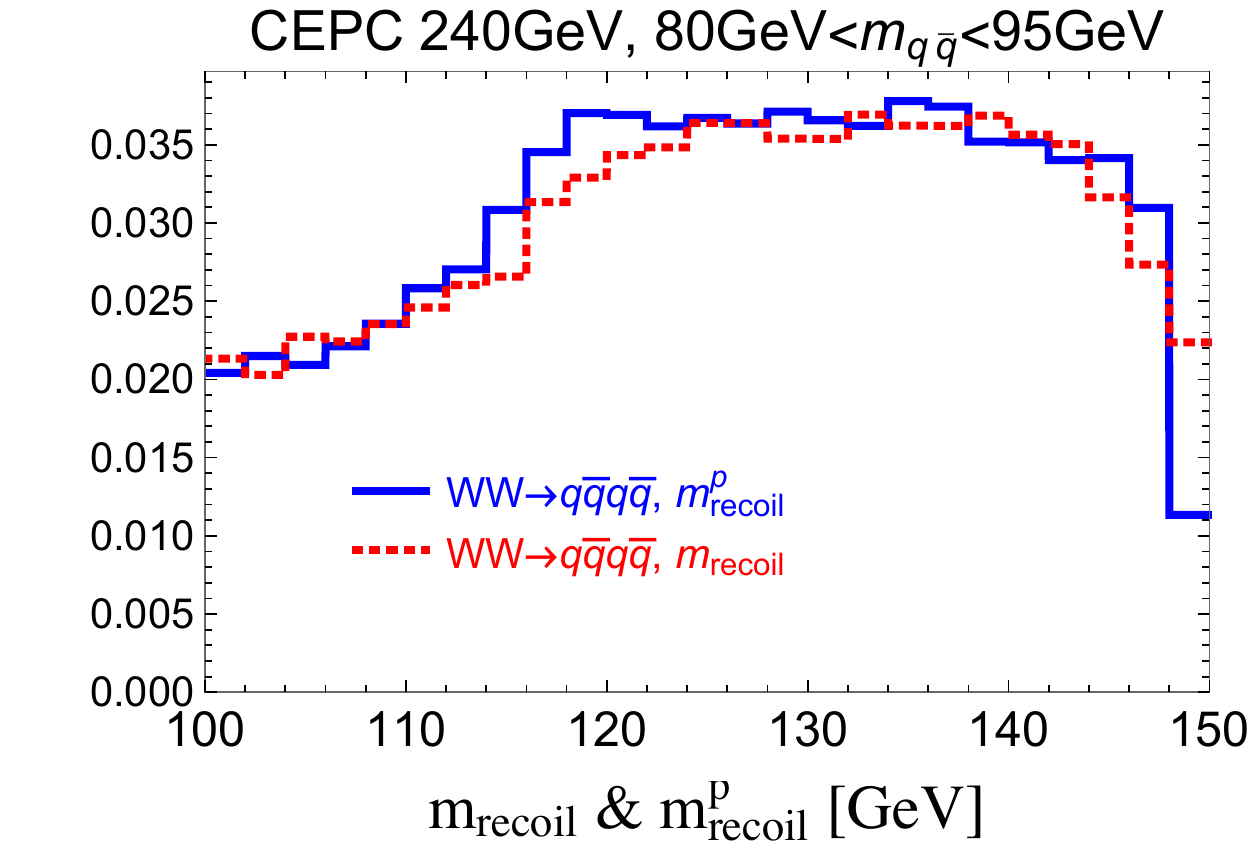}  
\caption{The distributions of  $\mrc$ and $\mrcp$ for $W^+W^-\rightarrow q\bar{q}q\bar{q}$ at CEPC 240\,GeV with a $Z$-mass-window cut  of $70\,{\rm GeV}<m_{q\bar{q}}<110\,{\rm GeV}$ (left panel) or $80\,{\rm GeV}<m_{q\bar{q}}<95\,{\rm GeV}$ (right panel).}
\label{fig:wwqqqq}
\end{figure}

Since the background events do not have Higgs in them, we do not expect the $\mrcp$ distribution of them to accumulate around the Higgs mass.  As a simple estimation, we show the $\mrc$ and $\mrcp$ distributions for one of the main backgrounds, $W^+W^-\rightarrow q\bar{q}q\bar{q}$ in Fig~\ref{fig:wwqqqq}, also for both choices of the $Z$-mass-window cuts.
It is interesting to notice that for the larger $Z$ mass window, $\mrcp$ actually reduces the background events in the region of $\sim$100-120\,GeV, while for the smaller window, the $\mrc$ and $\mrcp$ distributions are very similar. 

Our study shows that the $\mrcp$ variable could better reconstruct the Higgs mass for the signal and does not have the same effect on backgrounds. As such, we expect it to provide a significant improvement on the inclusive cross section measurements of the Higgsstrahlung process compared with the conventional recoil mass variable $\mrc$. Needless to say, such an improvement is crucially relevant to the studies of the Higgs boson properties.   We also find a similar behavior of the signal and background distributions at the ILC 250\,GeV, the results of which are not specifically shown.  Since we have only performed a simplified simulation analysis and have not considered some of the important backgrounds, we will restrain ourselves from doing any quantitative analysis on the inclusive $\sigma(hZ)$ measurements and leave it for the experimental groups who have better tools for such an analysis.


\section{Conclusions}
\label{sec:con}

In this paper, we explore the use of the recoil mass and its variations in the measurements of the $WW$-fusion process at a lepton collider with a center of mass energy of 240-250\,GeV.  We found the variable $\mrcp$, constructed using only the 3-momenta of the Higgs decay products, can better separate Higgsstrahlung events with an invisible $Z$ from the $WW$-fusion events than the original recoil mass $\mrc$ does, with an improvement up to 20-30\% on the precision of the $WW$-fusion cross section.  We study its impact in both the conventional framework and the effective-field-theory one.   In the conventional framework, a better precision on the $WW$-fusion cross section leads to a significant improvement on the constraints of the $hWW$ coupling  and the total Higgs width.  In a global analysis under the effective-field-theory framework, using the information in the $\mrc$ or $\mrcp$ distributions could improve the reach on some of the EFT parameters by more than $10\%$ compared with just using the total rate of the $\eevvh$ channel.  
We find that fitting the EFT parameters directly to the binned distributions gives the best precision reach.  On the other hand,  if the EFT parameters are fitted to the precisions of the $WW$-fusion and $hZ, Z\to \nu \bar{\nu}$ cross sections extracted from the $\mrc$ distribution, the precision reach could suffer from the large correlation between the two cross sections if it is not taken count of.  
We also explore the use of $\mrcp$ in the inclusive measurements of the Higgsstrahlung process ($\eehz$) with hadronic $Z$s and find that it can significantly improve the reconstruction of the Higgs at a center of mass energy of 240-250\,GeV.  The use of $\mrcp$ could therefore potentially lead to an improvement on the overall precision of the inclusive $hZ$ cross section measurements.  
The construction of $\mrcp$ is also extremely simple and does not require any additional measurements.  It should be straightforward to implement $\mrcp$ in any studies that make use of the recoil mass distribution. 
%


\subsection*{Acknowledgments}
We thank Tao Liu, Zhen Liu and Yan Wang for useful discussions and valuable comments on the manuscript.  JG is supported by an International Postdoctoral Exchange Fellowship Program between the Office of the National Administrative Committee of Postdoctoral Researchers of China (ONACPR) and DESY. YYL is supported by Hong Kong PhD Fellowship (HKPFS) and the Collaborative Research Fund (CRF) under Grant No HUKST4/CRF/13G.

\appendix


\section{The ``12 (or 11)-parameter'' effective-field-theory framework}
\label{app:eft}

We follow the framework in Ref.~\cite{Durieux:2017rsg} which uses the Higgs basis, proposed in Ref.~\cite{Falkowski:2001958} and applied also in the studies of LHC Higgs measurements in Refs.~\cite{Falkowski:2015fla, Falkowski:2015jaa}.  We focus on CP-even dimension-6 (D6) operators and omit the ones that induce fermion dipole interactions.  We also assume the $Z$-pole observables and $W$ mass to be SM-like, given that they are already very well constrained by LEP and can be further constrained with a $Z$-pole run at the future lepton colliders. 

The relevant parts in the Lagrangian of the SM and D6 operators are
\begin{equation}
\La \supset \La_{hVV} + \La_{hff} + \La_{\rm tgc}\,,  \label{eq:latot}
\end{equation}
where the Higgs boson couplings to a pair of SM gauge bosons are given by
\begin{align}
\La_{hVV} = &~ \frac{h}{v} \bigg[ (1+\delta c_W) \frac{g^2 v^2}{2} W^+_\mu W^{-\mu} + (1+\delta c_Z) \frac{(g^2+g'^2)v^2}{4} Z_\mu Z^\mu \nonumber\\
&~~~ + c_{WW} \, \frac{g^2}{2}W^+_{\mu\nu}W^{-\mu\nu} +c_{W\square}\, g^2(W^-_\mu \partial_\nu W^{+\mu\nu} +{\rm h.c.}) \nonumber\\
&~~~ + c_{gg} \, \frac{g^2_s}{4} G^a_{\mu\nu} G^{a\,\mu\nu} + c_{\gamma\gamma} \, \frac{e^2}{4} A_{\mu\nu} A^{\mu\nu} + c_{Z\gamma} \, \frac{e\sqrt{g^2+g'^2}}{2} Z_{\mu\nu} A^{\mu\nu} \nonumber\\
&~~~ + c_{ZZ} \, \frac{g^2+g'^2}{4} Z_{\mu\nu} Z^{\mu\nu} + c_{Z\square} \, g^2 Z_\mu \partial_\nu Z^{\mu\nu} + c_{\gamma\square} \, g g' Z_\mu \partial_\nu A^{\mu\nu} \, \bigg] \,. \label{eq:hvv}
\end{align}
The parameters in \autoref{eq:hvv} are not all independent.  Four constraints can be written down by imposing gauge invariances, which we choose to rewrite $\delta c_W$, $c_{WW}$, $c_{W\square}$ and $c_{\gamma\square}$ as 
\begin{align}
\delta c_W =&~ \delta c_Z + 4\delta m \,,  \nonumber\\
c_{WW} =&~ c_{ZZ} + 2s^2_{\theta_W} c_{Z\gamma} + s^4_{\theta_W} c_{\gamma\gamma} \,, \nonumber\\
c_{W\square} =&~ \frac{1}{g^2-g'^2} \left[  g^2 c_{Z\square} + g'^2 c_{ZZ} - e^2 s^2_{\theta_W} c_{\gamma\gamma} - (g^2-g'^2) s^2_{\theta_W} c_{Z\gamma}  \right]  \,, \nonumber \\
c_{\gamma\square} =&~ \frac{1}{g^2-g'^2} \left[  2 g^2 c_{Z\square} + (g^2+g'^2)c_{ZZ} - e^2 c_{\gamma\gamma} - (g^2-g'^2) c_{Z\gamma}  \right]  \,, \label{eq:cwtocz}
\end{align}
where $\delta m$ can only be induced by custodial symmetry breaking effects and is set to zero in our framework.  For the Yukawa couplings, we focus on the ones of $ t,\,c,\,b,\,\tau,\,\mu$, parameterized as
\begin{equation}
\La_{hff} = -\frac{h}{v} \sum_{f = t,c,b,\tau,\mu} m_{f} (1+\delta y_f ) \bar{f}_{R} f_{L} + {\rm h.c.} \,.  \label{eq:deltayf}
\end{equation}
The possible flavor violating Yukawa couplings from new physics are not considered.  The anomalous triple gauge couplings (aTGCs) are parameterized as 
\begin{align}
\La_{\rm tgc}  ~=~&~~~ i g s_{\theta_W}  A^\mu (W^{-\nu} W^+_{\mu\nu} - W^{+\nu} W^-_{\mu\nu}) \nonumber\\
&~ + ig (1+\delta g^Z_1) c_{\theta_W} Z^\mu (W^{-\nu} W^+_{\mu\nu} - W^{+\nu} W^-_{\mu\nu}) \nonumber\\
&~ + ig\left[ (1+\delta \kappa_Z) c_{\theta_W} Z^{\mu\nu} + (1+ \delta \kappa_\gamma) s_{\theta_W} A^{\mu\nu}\right] W^-_\mu W^+_\nu  \nonumber\\
&~ + \frac{ig}{m^2_W} (\lambda_Z c_{\theta_W} Z^{\mu\nu} + \lambda_\gamma s_{\theta_W} A^{\mu\nu}) W^{-\rho}_v W^+_{\rho\mu} \,, \label{eq:tgc}
\end{align}
where $V_{\mu\nu} \equiv \partial_\mu V_\nu - \partial_\nu V_\mu$ for $V = W^\pm,\, Z, \, A$.   Gauge invariance further imposes the relations $\delta \kappa_Z =\delta g_{1,Z} - t^2_{\theta_W} \delta \kappa_\gamma$ and $\lambda_Z =\lambda_\gamma$.  This leaves three independent aTGC parameters, which we choose to be $\delta g_{1,Z}$, $\delta \kappa_\gamma$ and $\lambda_Z$.  Two of them, $\delta g_{1,Z}$ and $\delta \kappa_\gamma$, are related to the Higgs parameters and can be written as
\begin{align}
\delta g_{1,Z} =&~ \frac{1}{2(g^2-g'^2)} \left[  -g^2(g^2+g'^2) c_{Z\square} -g'^2(g^2+g'^2)c_{ZZ} + e^2 g'^2 c_{\gamma\gamma} + g'^2(g^2-g'^2)c_{Z\gamma}  \right]  \,, \nonumber \\
\delta \kappa_\gamma =&~ -\frac{g^2}{2} \left(  c_{\gamma\gamma} \frac{e^2}{g^2+g'^2} + c_{Z\gamma}\frac{g^2-g'^2}{g^2+g'^2} - c_{ZZ}   \right) \, . \label{eq:tgchb}
\end{align}

To summarize, in our framework the contribution from D6 operators to the Lagrangian in \autoref{eq:latot} can be parametrized by the following 12 parameters:
\begin{equation}
	\delta c_Z		\,,~~
	c_{ZZ}			\,,~~
	c_{Z\square}		\,,~~
	c_{\gamma\gamma}	\,,~~
	c_{Z\gamma}		\,,~~
	c_{gg}			\,,~~
	\delta y_t		\,,~~
	\delta y_c		\,,~~
	\delta y_b		\,,~~
	\delta y_\tau		\,,~~
	\delta y_\mu		\,,~~
	\lambda_Z		\,.
\label{eq:para12}
\end{equation}

Also following Ref.~\cite{Durieux:2017rsg, Falkowski:2015fla, Falkowski:2015jaa}, we consider the EFT contribution to the $h\gamma\gamma$ and $hZ\gamma$ vertices at the tree level, in which case the only EFT parameter that contributes to the decay rate of $h\to \gamma\gamma$  ($h\to Z\gamma$) is $c_{\gamma\gamma}$ ($c_{Z\gamma}$).  For the decay $h\to gg$, we include, in addition to $c_{gg}$, the contributions of $\delta y_t$ and $\delta y_b$, which enter the $hgg$ vertex by modifying the Yukawa couplings in the fermion loops.  It is also convenient to normalize $c_{\gamma\gamma}$, $c_{Z\gamma}$ and $c_{gg}$ with respect to the SM 1-loop contributions.  We follow Ref.~\cite{Durieux:2017rsg} and define the following parameters
\begin{equation}
\frac{\Gamma_{\gamma\gamma}}{\Gamma^{\rm SM}_{\gamma\gamma}} 
	\simeq  1-2 \bar{c}_{\gamma\gamma}
	\,,  \hspace{1cm}
\frac{\Gamma_{Z\gamma}}{\Gamma^{\rm SM}_{Z\gamma}}
	\simeq 1-2 \bar{c}_{Z\gamma}
	\,,
\label{eq:barcvv}
\end{equation} 
and
\begin{equation}
\frac{\Gamma_{gg}}{\Gamma^{\rm SM}_{gg}} 
	~\simeq~
	1 + 2\bar{c}^{\rm \,eff}_{gg}
	~\simeq~
	1+ 2 \, \bar{c}_{gg} + 2.10 \, \delta y_t -0.10 \, \delta y_b
	\,,
\label{eq:barcgg}
\end{equation}
where $\bar{c}_{\gamma\gamma}$, $\bar{c}_{Z\gamma}$ and $\bar{c}_{gg}$ are related to the original parameters by
\begin{equation}
\bar{c}_{\gamma\gamma}	\simeq \frac{c_{\gamma\gamma}}{8.3\times 10^{-2}} \,, \hspace{1cm} 
\bar{c}_{Z\gamma}	\simeq \frac{c_{Z\gamma}}{5.9\times 10^{-2}} \,, \hspace{1cm}  
\bar{c}_{gg}		\simeq \frac{c_{gg}}{8.3\times 10^{-3}} \,.  \label{eq:cnorm}
\end{equation}
Furthermore, without measuring the $t\bar{t}h$ process at high energies ($\sqrt{s}\gtrsim 500\,$GeV) or at the LHC, the parameters $c_{gg}$ and $\delta y_t$ can not be independently constrained.  Since we focus on the 240-250\,GeV run at lepton colliders, we replace $c_{gg}$ and $\delta y_t$ by $\bar{c}^{\rm \,eff}_{gg}$ in \autoref{eq:barcgg} which parametrize the total contribution to the $hgg$ vertex.  The number of parameters is thus reduced to 11, and the parameters are
\begin{equation}
	\delta c_Z		\,,~~
	c_{ZZ}			\,,~~
	c_{Z\square}		\,,~~
	\bar{c}_{\gamma\gamma}	\,,~~
	\bar{c}_{Z\gamma}		\,,~~
	\bar{c}^{\rm \,eff}_{gg}			\,,~~
	\delta y_c		\,,~~
	\delta y_b		\,,~~
	\delta y_\tau		\,,~~
	\delta y_\mu		\,,~~
	\lambda_Z		\,,
\label{eq:para11}
\end{equation}
which are used in our EFT global analysis in \autoref{sec:heft}.  


\section{EFT expressions for the $\eevvh$ cross sections}
\label{app:expression}

We obtain the cross section of $\eevvh$ as a function of the EFT parameters by generating events using \texttt{Madgraph5}~\cite{Alwall:2014hca} with the \texttt{BSMC} package~\cite{BSMC, Falkowski:2015wza}.  The events are showered in \texttt{Pythia}~\cite{Sjostrand:2006za} and passed to \texttt{Delphes}~\cite{deFavereau:2013fsa} with the ILD card for detector simulations, after which the selection cuts in \autoref{sec:mrecoil} are applied.  The interference between $hZ$ and $WW$ fusion are also included.  The results for CEPC 240\,GeV with unpolarized beams are listed as follows.  For the total rate, we have 
\begin{align}
\left. \frac{\sigma_{\nu\bar{\nu}h}}{\sigma^{\rm SM}_{\nu\bar{\nu}h}} \right\vert_{ 240\,{\rm GeV}}^{\rm unpolarized} =&~ 
1 + 1.7 \, \delta c_Z + 1.3 \, c_{ZZ} + 2.9 \, c_{Z\square} + 0.051 \, c_{Z\gamma} +  0.14 \, c_{\gamma\square} \nonumber \\
&~ + 0.23 \, \delta c_W -0.0026 \, c_{WW}  -0.065  \, c_{W\square} \,.  \label{eq:rateeft}
\end{align}
Here we do not impose the gauge invariance condition from \autoref{eq:cwtocz} in order to show the different dependences on the $Z$ and $W$ parameters.
For the binned differential distributions of $\mrc$ ($\mrcp$), the numerical coefficients in \autoref{eq:rateeft} are replaced by the ones in \autoref{tab:csm} (\autoref{tab:csmp}).
\begin{table}[h]\scriptsize
\centering
\begin{tabular}{|c|cccccccccc|} \hline
\multicolumn{11}{|c|}{CEPC 240\,GeV (with unpolarized beams) $m_{recoil}$}  \\  \hline
&   \multicolumn{10}{c|}{bin index [GeV]}\\  
 & 75 & 80 & 85 & 90 & 95 & 100 & 105 & 110 & 115 & 130 \\ \hline
$ \sigma_{SM}$[fb] & 0.15 & 0.18 & 0.38 & 0.78 & 1.2 & 1.3 & 1.1 & 0.74 & 0.47 & 0.34 \\ \hline
 $\delta c_Z$ & 0.97 & 1.4 & 1.6 & 1.7 & 1.8 & 1.9 & 1.9 & 1.9 & 1.8 & 1.9 \\
 $c_{ZZ}$ & 0.50 & 0.95 & 1.1 & 1.3 & 1.3 & 1.4 & 1.4 & 1.4 & 1.4 & 1.4 \\
 $c_{Z\square}$ & 1.5 & 2.2 & 2.6 & 2.9 & 3.0 & 3.0 & 3.2 & 3.1 & 3.1 & 3.0 \\
 $c_{Z\gamma}$ & 0.021 & 0.035 & 0.044 & 0.051 & 0.052 & 0.054 & 0.056 & 0.056 & 0.055 & 0.055 \\
 $c_{\gamma\square}$ & 0.075 & 0.11 & 0.13 & 0.14 & 0.15 & 0.15 & 0.15 & 0.15 & 0.15 & 0.15 \\
 $\delta c_W$ & 0.93 & 0.62 & 0.37 & 0.22 & 0.15 & 0.13 & 0.14 & 0.17 & 0.20 & 0.29 \\
 $c_{WW}$ & -0.011 & -0.0066 & -0.0038 & -0.0023 & -0.0016 & -0.0013 & -0.0013 & -0.0016 & -0.0019 & -0.0023 \\
 $c_{W\square}$ & -0.30 & -0.18 & -0.11 & -0.060 & -0.038 & -0.032 & -0.033 & -0.036 & -0.037 & -0.049 \\
 \hline
\end{tabular}
\caption{The coefficients of the EFT parameters for the expression of $\sigma/\sigma_{\rm SM}$ of each bin of the $\mrc$ distribution.  The upper bound of each bin is listed in the first row.  The first bin include all the events below 75\,GeV.   
}
\label{tab:csm}
\end{table}
\begin{table}[h]\scriptsize
\centering
\begin{tabular}{|c|cccccccc|} \hline
\multicolumn{9}{|c|}{CEPC 240\,GeV (with unpolarized beams) $m^p_{recoil}$}  \\  \hline
&   \multicolumn{8}{c|}{bin index  [GeV]}\\ 
 & 75 & 80 & 85 & 90 & 95 & 100 & 105 & 115 \\ \hline
 $\sigma_{SM}$[fb] & 0.22 & 0.24 & 0.59 & 1.7 & 2.4 & 0.99 & 0.32 & 0.11 \\ \hline
 $\delta c_Z$ & 0.95 & 1.4 & 1.7 & 1.8 & 1.9 & 2.0 & 1.8 & 1.4 \\
 $c_{ZZ}$ & 0.54 & 0.99 & 1.2 & 1.3 & 1.4 & 1.4 & 1.4 & 1.0 \\
 $c_{Z\square}$ & 1.6 & 2.3 & 2.6 & 3.0 & 3.2 & 3.3 & 3.1 & 2.0 \\
 $c_{Z\gamma}$ & 0.021 & 0.035 & 0.045 & 0.053 & 0.056 & 0.059 & 0.054 & 0.044 \\
 $c_{\gamma\square}$ & 0.075 & 0.11 & 0.13 & 0.15 & 0.16 & 0.16 & 0.15 & 0.11 \\
 $\delta c_W$ & 0.92 & 0.61 & 0.33 & 0.14 & 0.075 & 0.11 & 0.33 & 0.99 \\
 $c_{WW}$ & -0.0095 & -0.0062 & -0.0034 & -0.0014 & -0.00082 & -0.0012 & -0.0025 & -0.0075 \\
 $c_{W\square}$ & -0.28 & -0.17 & -0.092 & -0.037 & -0.019 & -0.025 & -0.056 & -0.13 \\ \hline
\end{tabular}
\caption{Same as \autoref{tab:csm} but for the $\mrcp$ distribution.}
\label{tab:csmp}
\end{table}
We have also checked that the statistical uncertainties from simulation are under control.\footnote{There are nevertheless some small fluctuations in our result.  For instance, the coefficients of $\delta c_Z$ and $\delta c_W$ should always add up to two.  In most bins, the sum is controlled in the range 1.9--2.1.  After imposing $\delta c_Z = \delta c_W$ we simply fix its coefficient to 2.  We do not expect the fluctuations in other coefficients to significantly change our results.}  We then impose the gauge invariance condition in \autoref{eq:cwtocz} and construct $\chi^2$ of the $\eevvh$ measurement, assuming the events follow a poisson distribution.  For the total rate, we have 
\begin{equation}
\chi^2 = \frac{N^2_{\rm sig} \left(1-\frac{\sigma_{\nu\bar{\nu}h}}{\sigma^{\rm SM}_{\nu\bar{\nu}h}}\right)^2 }{ N_{\rm sig} + N_{\rm bg} } \,,  \label{eq:chis}
\end{equation}
where $N_{\rm sig}$ and $N_{\rm bg}$ are the number of signal and backgrounds after cuts, normalized to $5\inab$ for CEPC.  For the binned distributions, we use \autoref{eq:chis} to construct the $\chi^2$ of each bin where $N_{\rm sig}$ and $N_{\rm bg}$ are the number of signal and backgrounds in the bin. We then sum over the $\chi^2$ of all the bins, assuming no correlation among them.
The $\chi^2$ is then combined with the ones of other measurements for the global analysis in \autoref{sec:heft}.
We refer the readers to Ref.~\cite{Durieux:2017rsg} for a complete set of expressions for the other relevant observables.


\providecommand{\href}[2]{#2}\begingroup\raggedright\endgroup


\end{document}